\documentclass[12pt,letterpaper]{article}

\usepackage{geometry}
\usepackage{setspace}
\spacing{1.2}

\usepackage[utf8]{inputenc}
\usepackage[style=chem-acs, articletitle=true]{biblatex}
\let\oldcite\cite
\renewcommand{\cite}[1]{\textsuperscript{\oldcite{#1}}}

\usepackage{amsmath}
\usepackage{amssymb}
\usepackage{amsfonts}
\usepackage{float} 

\usepackage{longtable}
\usepackage{booktabs}
\usepackage{pdflscape}
\usepackage{array}
\usepackage{cellspace}

\usepackage{tablefootnote} 
\usepackage{graphicx}
\usepackage{dcolumn}
\usepackage{bm}
\usepackage{xfrac}
\usepackage[utf8]{inputenc}
\usepackage[T1]{fontenc}
\usepackage{mathptmx}
\usepackage{etoolbox}
\usepackage{xcolor}

\numberwithin{equation}{section}

\usepackage{mathrsfs}

\usepackage{multirow}
\usepackage{enumitem} 

\usepackage{xr}
\usepackage{comment}

\renewcommand{\arraystretch}{1.5}

\newcommand{\tp}{t^\prime}
\newcommand{\Jp}{J^\prime}
\newcommand{\Gp}{G^\prime}
\newcommand{\Jpp}{J^{\prime\prime}}
\newcommand{\Gpp}{G^{\prime \prime}}

\newcommand{\omg}{(\omega)}
\newcommand{\Zp}{Z^\prime}
\newcommand{\Zpp}{Z^{\prime\prime}}
\newcommand{\Zstar}{Z^*}

\newcommand{\epsstar}{\epsilon^*}
\newcommand{\epsp}{\epsilon^\prime}
\newcommand{\epspp}{\epsilon^{\prime \prime}}

\newcommand{\hbt}[1]{\mathcal{H}{\left[ #1 \right]}}

\newcommand{\review}[1]{\textcolor{black}{#1}}
\newcommand{\rereview}[1]{\textcolor{black}{#1}}
\newcommand{\rrereview}[1]{\textcolor{black}{#1}}

\newcommand{\addition}[1]{\textcolor{black}{#1}}

\usepackage{authblk}
\author[1]{Shivangi Mittal}
\author[2]{Sachin Shanbhag*}
\author[1]{Yogesh M. Joshi*}
\affil[1]{Department of Chemical Engineering, Indian Institute of Technology Kanpur, Kanpur - 208016, INDIA}
\affil[2]{Department of Scientific Computing, Florida State University, Tallahassee, FL 32306. USA}

\title{When Spectroscopies Speak the Same Language: Unifying Rheology, Electrochemical Impedance, and Dielectrics}
\date{* Email: sshanbhag@fsu.edu, joshi@iitk.ac.in}

\begin{document}

\maketitle

\begin{abstract}
    Spectroscopic techniques measure the dynamical response of physical systems subjected to oscillatory perturbations. \review{For small perturbations around the equilibrium state, these spectroscopic methods are unified by the common mathematical framework of linear response theory.} This work presents a unified perspective on rheological or mechanical spectroscopy, electrochemical impedance spectroscopy, and broadband dielectric spectroscopy through the lens of linear response theory. Subtle conceptual similarities and differences among these techniques are highlighted by analyzing their mapping to the linear response theory, basic building elements, elementary models, and time and frequency domain response functions. Data validation and analysis approaches, including Kramers–Kronig relations, equivalent circuits, and multimode models are discussed. The shared fundamentals of different spectroscopies enable seamless exchange of ideas across domains.

\end{abstract}

\section*{Keywords}
Linear response, rheology, electrochemical impedance spectroscopy, broadband dielectric spectroscopy, elementary models, time and frequency domain response

\clearpage




\section{Introduction}
\label{sec:intro}

Spectroscopy is the study of how matter interacts with energy -- most commonly electromagnetic radiation.\cite{hofmann2010spectroscopic}  \review{However, the term spectroscopy may be used in a broader context, encompassing studies that investigate the dynamical response of a system to an oscillatory stimulus by analyzing the absorption, emission, and scattering of energy across a spectrum of frequencies or timescales.\cite{magalas2003mechanical} }
The nature of the external perturbation ranges from light,\cite{parson2007modern} electric,\cite{kremer2002broadband, Lasia2014} magnetic,\cite{jackman2012dynamic} mechanical,\cite{magalas2003mechanical, hecksher2017toward} acoustic,\cite{Dukhin2000127} etc. Spectroscopy seeks to disentangle processes that occur at the same \textit{time} but at different \textit{timescales} or frequencies like a prism that separates visible light into its constituent colors or frequencies. Although different spectroscopic techniques probe different aspects of materials, they are animated by a common conceptual and mathematical framework.  

The goal of this primer is to provide a gentle mathematical introduction to this framework through the lens of mechanical, electrochemical impedance, and broadband dielectric spectroscopies. Despite the common conceptual bedrock of the linear response theory (LRT), there are subtle but sometimes confusing differences in conventions, elementary models, and methods of analysis. Therefore, a secondary goal of this paper is to cast light on these differences. The intended audience is graduate students or practitioners familiar with one of the three methods.
\begin{enumerate}[label=(\roman*)]

\item \textit{Mechanical spectroscopy} (MS) studies how energy is stored or dissipated when a material is subjected to a time-dependent mechanical perturbation.\cite{magalas2003mechanical} This perturbation typically takes the form of an oscillatory strain $\gamma$ (or stress $\sigma$) field. The resulting stress (or strain) field is measured. Stress $(\sigma)$ and strain ($\gamma$) are \textit{conjugate variables}: they form a pair of stimulus and response variables whose roles may be reversed. The history of systematic experiments using oscillatory methods to measure the elasticity of materials can be traced to the early 1900s.\cite{Poynting1909} In MS, the accessible frequency window in routine laboratory measurements typically spans only a few decades (\review{$10^{-3}$--$10^{2}$ Hz}), despite modern rheometers claiming access to much lower values ($\sim 10^{-6}$ \review{Hz}).  MS subsumes dynamic mechanical analysis in which tension, compression, bending, or torsional excitation is applied to solid/semi-solid materials like metals, composites, glasses etc.\cite{Menard2020} MS also includes shear rheology, common for softer materials like gels, suspensions, polymer solutions and melts, etc. For brevity, we limit our discussion to the realm of shear rheology, although extensions to dynamic mechanical analysis are straightforward.\cite{Ferry1980} 


\item \textit{Electrochemical impedance spectroscopy} (EIS) is a non-destructive characterization technique used to identify transport mechanisms in electrochemical systems, such as batteries, fuel cells, corrosion and coating, supercapacitors, sensors, and electrochemical interfaces in general.\cite{wang2021electrochemical, Lasia2014} Reaction kinetics at an electrode/electrolyte interface are governed by the interplay of transport phenomena with slow and fast dynamics, such as charge transfer, mass transfer, and adsorption. The electrochemical system is hooked to a circuit and an alternating voltage (or current) is applied, while the resulting current (or voltage) is measured. The conjugate variables in EIS are voltage ($V$) and electric charge ($Q$). The latter is related to the current $I$ via $I = \dot{Q}$ (the dot indicates a time derivative). Although commercial EIS analyzers often advertise frequency ranges spanning from $\mu$Hz to MHz, the practical lower limit is $\sim 1$ mHz.\cite{Lazanas2023}


\item \textit{Broadband dielectric spectroscopy} (BDS) belongs to the class of impedance spectroscopy in which an alternating voltage (current) is applied to a parallel plate capacitor containing the dielectric material of interest, and the resulting current (voltage) is measured.
\rereview{The response is intrinsically linked to the total current density consisting of the displacement current arising from polarization of bound charges under the field, and the conduction current due to transport of mobile charges.} 
The conjugate variables in BDS are the electric field ($E$) and the dielectric displacement ($D$), which are related to polarization. BDS captures an exceptionally wide frequency window of nearly 18 decades ($10^{-6} - 10^{12}$ Hz).\cite{woodward2021broadband} 
This broad range enables the investigation of dynamic phenomena spanning decades, capturing both rapid local vibrations and slow cooperative motions.\cite{kremer2002broadband} BDS is applied to a plethora of systems from soft materials, ionic conductors, biomaterials, composites, etc.\cite{Hedvig1977, el2016dielectric, kao2004dielectric, havriliak1997dielectric, woodward2021broadband}


\end{enumerate}

Table \ref{tab:x_y} summarizes the frequency range and conjugate variables for these three spectroscopies. For concreteness, we shall hereon identify the conjugate variable $x$ with a generalized ``displacement'' and the conjugate variable $y$  with a generalized ``force'' or ``potential''.

\begin{table}
    \centering
\begin{tabular}{lllll}
\hline
& range (Hz) & force $y(t)$ & displacement $x(t)$ & velocity $\dot{x}(t)$ \\
\hline 
MS & $10^{-6} - 10^{2}$& stress $\sigma$ & strain $\gamma$ & strain rate $\dot{\gamma}$ \\
EIS & $10^{-6} - 10^{6}$ & voltage $V$ & charge $Q$ & current $I$ \\
BDS & $10^{-6} - 10^{12}$  & electric field $E$ & dielectric displacement $D/\epsilon_0$ & current density $j/\epsilon_0$ \\
\hline
\end{tabular}
    \caption{The frequency range and conjugate variables (generalized force $y(t)$ and displacement $x(t)$) in MS, EIS, and BDS. The last column shows the generalized velocity, where the dot indicates a time derivative. In the last row, $\epsilon_0$ is the permittivity of vacuum.}
    \label{tab:x_y}
\end{table}

\subsection{Scope and Layout}

We begin by introducing the common theoretical framework that underpins all three techniques: the linear response theory or LRT (section \ref{sec:lrt}). Subsequently, we separately describe domain-specific terminology and elementary models of MS, EIS, and BDS (sections \ref{sec:ms} -- \ref{sec:bds}) before comparing and contrasting them in section \ref{sec:discussion}. The scope of this primer is limited by design; readers are directed to authoritative texts and reviews on MS,\cite{Ferry1980, tschoegl2012phenomenological, cho2016viscoelasticity, shaw2018introduction} EIS,\cite{Lasia2014, orazem2020tutorial, wang2021electrochemical, Lazanas2023} and BDS \cite{havriliak1997dielectric, jonscher1999dielectric, kremer2002broadband, woodward2021broadband, venkatesh2005overview} for more comprehensive treatments.

\section{Linear Response Theory}
\label{sec:lrt}

\begin{figure}
\begin{center}
\includegraphics[width=0.7\textwidth]{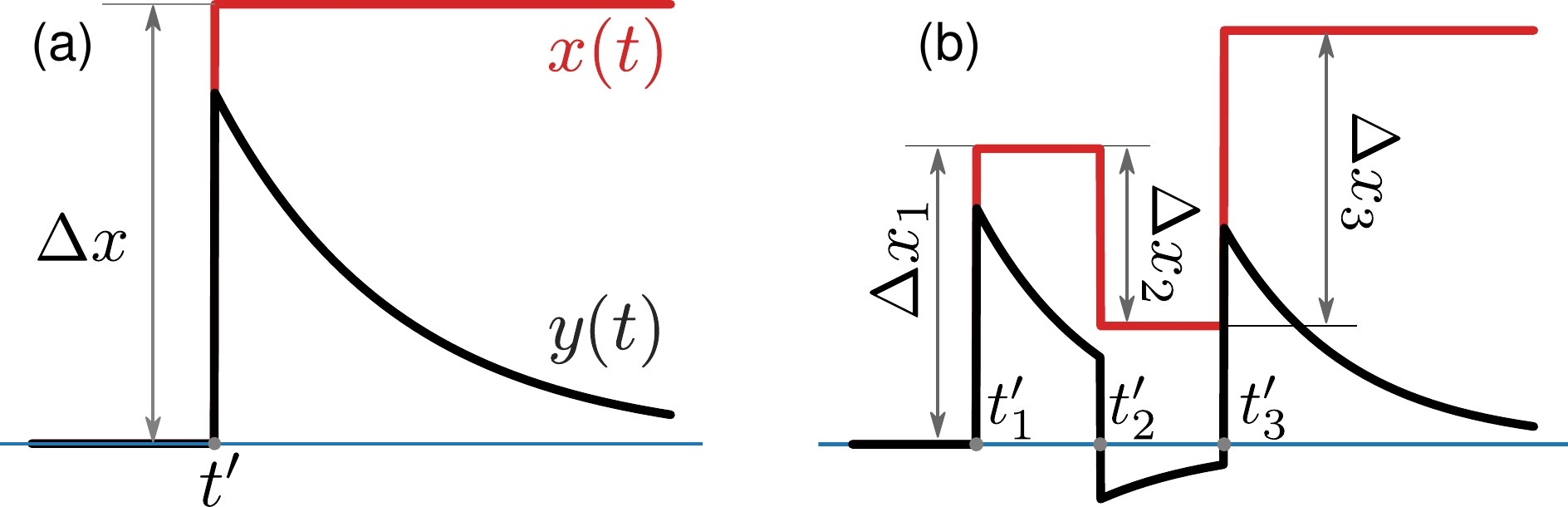}
\end{center}
\caption{The response $y(t)$ according to the LRT \addition{(section \ref{sec:lrt})} with an exponential response function $R_x(v) = e^{-v}$ to (a) a step perturbation $x(t \geq t^{\prime}) = \Delta x$, (b) three sequential excitations at different times as indicated.
\label{fig:lrt}}
\end{figure}

Close to equilibrium, the response of a dynamical system is approximately linear. This means that when a \textit{tiny} displacement $\Delta x$ is applied at time $\tp$ to a system at equilibrium, it produces a proportional change $\Delta y$ in the conjugate or force variable. LRT describes the evolution of $\Delta y$, shown schematically in figure \ref{fig:lrt}a, via a \textit{response function} $R_{x}(t - \tp)$ that only depends on the time elapsed since perturbation, $t - \tp$. This property is called \textit{time-invariance} and may be expressed as:
\begin{equation}
\Delta y(t) = y(t) - y_\text{eq} = R_{x}(t - \tp)\, \Delta x(\tp).
\label{eqn:basic}
\end{equation}
 Here $y_\text{eq}$ is the equilibrium value of $y$, which we shall set to zero for simplicity. Response functions obey another important property called \textit{causality}, $R_{x}(t - \tp) = 0$ for $t < \tp$, which mathematically encodes the idea that cause precedes effects. Therefore, \review{for $v=t-\tp$}, $R_{x}(v)$ takes a single argument, and is nonzero only when $v \geq 0$.  

When $n$ small perturbations $\Delta x_i$ are applied to the system at times $\tp_i$ where $1 \leq i \leq n$, linearity implies that the total response is given by the linear superposition (see figure \ref{fig:lrt}b) of the individual responses:
\begin{equation}
y(t) = \sum_{i=1}^{n} \Delta y_i(t) = \sum_{i=1}^{n} R_{x}(t - \tp_i) \, \Delta x(\tp_i).
\end{equation}
In the limit of continuous deformation this becomes
\begin{equation}
    y(t) = \int_{-\infty}^{t} R_{x}(t - \tp) \dot{x}(\tp) \, d\tp \label{eqn:lrt}
\end{equation} 
and encapsulates the LRT. It may be recast into alternative forms by recognizing that \review{eqn.} \eqref{eqn:lrt} describes a \textit{convolution} between $R_x$\addition{, a causal function,} and $\dot{x} = dx(\tp)/d\tp$,
\begin{equation}
y(t) = (R_{x} * \dot{x})(t) = \int_{-\infty}^{\infty} R_{x}(t - \tp) \, \dot{x}(\tp)  d\tp = \int_{0}^{\infty} R_{x}(\tp) \dot{x}(t-\tp) \, d\tp.
\label{eqn:lrt-conv}
\end{equation}
See \rrereview{section S3.1 in the supplementary material} for the relationship between causality and the limits of integration.  
Integrating the last expression in eqn \eqref{eqn:lrt-conv} by parts yields another useful form
\begin{equation}
y(t) =  R_{x}(0) x(t) + \int_{0}^{\infty} \dot{R}_x (\tp) x(t-\tp) d\tp.
\label{eqn:lrt2}
\end{equation}
\review{For eqns. \eqref{eqn:lrt} and \eqref{eqn:lrt-conv} to hold, $\dot{R}_x(t) \leq 0$, i.e. $R_x(t)$ is monotonic and does not increase with $t$.}

A useful attribute of response functions is the \textit{impulse property}. Consider the response to a step jump in the input $x(t) = \Delta x\, H(t)$, where $H(t)$ is the Heaviside step function. The derivative $\dot{x} = \Delta x \, \delta(t)$ describes an impulse, where $\delta(\cdot)$ is the Dirac delta function. From \review{eqn.} \eqref{eqn:lrt-conv}
\begin{equation}
y(t)  = \int_{-\infty}^{\infty} R_{x}(t - \tp)\, \Delta x\, \delta(\tp)\, d\tp = R_x(t) \Delta x.
\label{eqn:impulse_response}
\end{equation}
Thus, the response function $R_x(t) = y(t)/\Delta x$ can be determined by measuring the outcome of a step input jump. 

In LRT, the role of the stimulus and response variables can be switched due to their conjugate relationship. In other words, when we apply a perturbation $\Delta y(\tp)$ to a system at equilibrium, the resulting displacement $\Delta x(t) = x(t) - x_\text{eq} = R_y(t-\tp) \Delta y(\tp)$. Assuming $x_\text{eq} = 0$, it follows that
\begin{align}
x(t) & = (R_{y} * \dot{y})(t) = \int_{0}^{\infty} R_{y}(\tp) \dot{y}(t-\tp) \,  d\tp, \label{eqn:lrt_reverse0}
\\ 
&=  R_{y}(0) y(t) + \int_{0}^{\infty} \dot{R}_{y}(\tp)\,  y(t-\tp) \, d\tp.
\label{eqn:lrt_reverse}
\end{align} 
The physical units of the response functions $R_x$ and $R_y$ are the inverse of each other, but $R_x \neq 1/R_y$. 
\review{Theoretically, the LRT is valid only for vanishingly small inputs. In practice, however, nonlinear effects are typically negligible for small but finite perturbations. Furthermore, nonlinear spectroscopy --- wherein large perturbations are intentionally applied to quantify nonlinear effects --- is an important technique in its own right, but is beyond the scope of this work.}

Since $x$ and $y$ can be thought of as generalized displacement and force, respectively, $R_x$ represents a ``generalized modulus'' that is analogous to stiffness. For a perturbed system showing a time-dependent response, $R_x$ also tells us something about how the material dissipates energy as it relaxes back to equilibrium. Similarly, $R_y$ is a ``generalized compliance'' which is indicative of the ease with which a material can be deformed by applying force. The mathematical relationship between $R_x$ and $R_y$ can be elucidated by considering Laplace transforms. Please see section S1 in the supplementary material for a primer on Laplace transforms and how they relate to the LRT.

\subsection{Relation between Response Functions}

The relation between the response functions $R_x$ and $R_y$ is more transparent in the Laplace domain (indicated by tilde) where convolutions take a simpler form. \review{The Laplace transform of eqns. \eqref{eqn:lrt-conv} and \eqref{eqn:lrt_reverse0} yields,}
\begin{gather}
    \tilde{y}(s) = \tilde{R}_{x}(s) \tilde{\dot{x}}(s) = s \tilde{R}_{x}(s) \tilde{x}(s).
\label{eqn:lrt1_laplace} \\
\tilde{x}(s) = \tilde{R}_{y}(s) \tilde{\dot{y}}(s) = s \tilde{R}_{y}(s) \tilde{y}(s),
\label{eqn:lrt_laplace_reverse}
\end{gather}
which leads to the convolution relation in the Laplace domain,
\begin{equation}
\tilde{R}_{x}(s)\, \tilde{R}_{y}(s) = \dfrac{1}{s^2}.
\label{eqn:convolution_laplace}
\end{equation}
The inverse Laplace transform of both sides yields the convolution relation in the time-domain,
\begin{equation}
\int_{0}^{t} R_{x}(\tp) \, R_{y}(t-\tp)\,d\tp = t.
\label{eqn:convolution_time}
\end{equation}
These relations are useful for interconversion: if either $R_x$ or $R_y$ is known, the other can be mathematically or numerically determined.

The key objective of MS, EIS, and BDS is to figure out $R_x$ and $R_y$. Theoretically, this can be determined by a step input due to the impulse property of response functions. In practice, however, applying an oscillatory perturbation is often preferred because it avoids artifacts that arise from sudden jumps and sharp ramps.

\subsection{Response to Oscillatory Stimulation}

Spectroscopy typically involves studying the response of a system subjected to periodic excitation over a spectrum of frequencies or timescales. Fourier transforms (and special cases like sine and cosine transforms) provide a natural language to describe periodic signals and are briefly reviewed in the supplementary material section S2. They are denoted by a hat (e.g. $\hat{x}$).

Consider the response $y(t)$ to a sinusoidal stimulus  $x(t) = x_0 \sin \omega t$ with $x_0 \ll 1$. Starting with the LRT (eqn. \eqref{eqn:lrt-conv}), using the cosine subtraction formula and invoking the sine and cosine transforms defined as
\begin{align}
\hat{R}_x^\text{sin} (\omega) & = \int_{0}^{\infty} R_x(v) \sin \omega v \, dv && \text{sine transform}\\
\hat{R}_x^\text{cos} (\omega) & = \int_{0}^{\infty} R_x(v) \cos \omega v \, dv && \text{cosine transform},
\end{align}
we find that the response
\begin{equation}
y(t) =  \int_{0}^{\infty} R_x(v)\, x_0 \omega \cos \left( \omega (t-v) \right) \, dv =  x_0 \omega \left[\hat{R}_x^\text{sin} (\omega) \sin \omega t + \hat{R}_x^\text{cos}(\omega) \cos \omega t \right],
\label{eqn:oscillatory_lrt}
\end{equation}
is also sinusoidal and consists of one component that is in-phase with the input perturbation, while the other is out-of-phase. The Fourier transform of the response function is related to the sine and cosine functions via $\hat{R}_x=\hat{R}_x^\text{cos}-i\hat{R}_x^\text{sin}$. 

Since the Fourier transform of a causal function is a special case of the Laplace transform (section S2), the convolution relations \review(eqns. \eqref{eqn:lrt1_laplace}-\eqref{eqn:convolution_laplace}) follow with $s=i\omega$,
\begin{equation}
    \hat{y}(\omega)  = i \omega \hat{R}_{x}(\omega) \hat{x}(\omega), \quad
    \hat{x}(\omega)  = i \omega \hat{R}_{y}(\omega) \hat{y}(\omega), \quad \text{and} \quad
    \hat{R}_{x}(\omega)\, \hat{R}_{y}(\omega)  = -\omega^{-2} 
    \label{eqn:lrt_ft}
\end{equation}

Interestingly, the amount of energy stored and dissipated over one cycle is related to $\hat{R}_x^\text{cos}$ and $\hat{R}_x^\text{sin}$, respectively. Since $y$ and $x$ represent generalized force and displacement, respectively, the work done is $W = \int y\,dx = \int y(t) \, \dot{x}(t) \, dt$. Therefore, the average work done over one period $T = 2\pi/\omega$
\begin{equation}
W = \int_{0}^{T} y(t)\, \dot{x}(t) \,dt = x_0^2 \omega^2 \int_0^{T} \left[\hat{R}_x^\text{sin}(\omega) \sin{\omega t} + \hat{R}_x^\text{cos}(\omega) \cos{\omega t}  \right]  \cos{\omega t}\, dt.
\label{eqn:osc_work_yx_1}
\end{equation}
We can partition the total work $W = W^{S} + W^{D}$ into the energy stored in the system as potential energy $(W^{S})$, and the energy lost due to irreversible mechanisms $(W^{D})$. The component of $y(t)$ that is in-phase with $x(t)$ contributes to $W^{S}$, while the out-of-phase one contributes to $W^{D}$. Consequently,
\begin{equation}
W^{S}=x_0^2 \omega^2 \hat{R}_x^\text{sin}\omg \int_0^T \sin{\omega t} \cos{\omega t} dt = 0. \label{eqn:energy_storage}
\end{equation}
implying that no net energy is stored over a complete cycle. On the other hand,
\begin{equation}
W^{D}= x_0^2 \omega^2 \hat{R}_x^\text{cos}\omg \int_0^T \cos^{2}{\omega t} dt = \pi \omega x_0^2 \, \hat{R}_x^\text{cos}(\omega). \label{eqn:energy_dissipate}
\end{equation}
\review{indicating a net energy dissipation quantified by the cosine transform of $R_x$.}

Table \ref{tab:lrt_relations} summarizes the time and frequency domain LRT relations for MS, EIS, and BDS as they typically appear in the literature: the precise correspondence with LRT is discussed separately in subsequent sections. \rrereview{Symbols and notations used in MS, EIS, and BDS are summarized in tables \ref{tab:symbols_ms}, \ref{tab:symbols_eis}, and \ref{tab:symbols_bds}, respectively.} Since EIS is primarily a frequency-domain technique, time-domain counterparts analogous to MS and BDS are seldom used.

\begin{landscape}
\setlength{\cellspacetoplimit}{8pt}
\setlength{\cellspacebottomlimit}{8pt}
\setlength{\LTcapwidth}{\linewidth}
\begin{longtable}{Sl|ScScSc}  
\caption{Mapping of LRT relations with MS, EIS, and BDS. EIS is inherently a frequency-domain technique. \label{tab:lrt_relations} } \\
\hline
\textbf{Relation} & \textbf{MS} & \textbf{EIS} & \textbf{BDS}  \\
\hline
\endfirsthead
\hline
\textbf{Relation} & \textbf{MS} & \textbf{EIS} & \textbf{BDS}  \\
\hline
\endhead
\hline
\multicolumn{4}{r}{\textit{Continued on next page}} \\
\endfoot
\hline
\endlastfoot
\textbf{Linear response} & Boltzmann Superposition & & \\[-2ex]
$\displaystyle y(t) = \int_{-\infty}^{t} R_{x}(t - \tp ) \dot{x}(\tp) \, d\tp $ 
&
$\displaystyle \sigma(t) = \int_{-\infty}^{t} G(t - \tp) \dot{\gamma}(\tp) \, d\tp $ &  &
$\displaystyle E(t) = \epsilon_0^{-1} \int_{-\infty}^{t} M(t - \tp) \dot{D}(\tp) \, d\tp $
\\[-1ex]
$\displaystyle x(t) = \int_{-\infty}^{t} R_{y}(t - \tp) \dot{y}(\tp) \, d\tp $ 
&
$ \displaystyle \gamma(t) = \int_{-\infty}^{t} J(t - \tp) \dot{\sigma}(\tp) \, d\tp $ & &
$\displaystyle D(t) = \epsilon_0 \int_{-\infty}^{t} \epsilon(t - \tp) \dot{E}(\tp) \, d\tp $
\\
\hline 
\textbf{Convolution Relation} & & & \\[-2ex]
$\displaystyle \int_{0}^{t} R_{x}(\tp) \, R_{y}(t-\tp)\,d\tp = t$
&
$\displaystyle \int_{0}^{t} G(\tp) \, J(t-\tp)\,d\tp = t$ 
& &
$\displaystyle \int_{0}^{t} M(\tp) \, \epsilon(t-\tp)\,d\tp = t$ 
\\
\hline
\textbf{Fourier Transform} & & & \\[-2ex]
$\displaystyle \hat{R}_x (\omega) = \int_{0}^\infty R_x (t) \, e^{-i \omega t} dt $
&
$\displaystyle  \dfrac{G^*(\omega)}{i\omega} = \int_{0}^\infty G(t) \, e^{-i \omega t} dt $
& &
$\displaystyle  \dfrac{M^*(\omega)}{i\omega} = \int_{0}^\infty M(t) \, e^{-i \omega t} dt $
\\ [-1ex]
$\displaystyle \hat{R}_y (\omega) = \int_{0}^\infty R_y (t) \, e^{-i \omega t} dt $
&
$\displaystyle  \dfrac{J^*(\omega)}{i\omega} = \int_{0}^\infty J(t) \, e^{-i \omega t} dt $
& &
$\displaystyle  \dfrac{\epsilon^*(\omega)}{i\omega} = \int_{0}^\infty \epsilon(t) \, e^{-i \omega t} dt $
\\
\hline
\textbf{Frequency Response} & & & \\[-4ex]
$\displaystyle \hat{y}(\omega) = i\omega \hat{R}_x (\omega) \, \hat{x}(\omega)$
&
$\displaystyle \hat{\sigma}(\omega) = G^* (\omega) \, \hat{\gamma}(\omega)$ &
$\displaystyle \begin{aligned}
    \hat{V}(\omega) & = i\omega Z^* (\omega) \, \hat{Q}(\omega) \\
    & = Z^* (\omega) \, \hat{I}(\omega)
\end{aligned}
$ &
$\displaystyle \begin{aligned}
    \hat{E} (\omega) &= M^*(\omega) \,  \hat{D}(\omega)/\epsilon_0 \\
    &= \sigma^*(\omega) \,  \dfrac{\hat{j}(\omega)}{i\omega}
\end{aligned}
$ 
\\[-2ex]
$\displaystyle \hat{x}(\omega) = i\omega \hat{R}_y (\omega) \, \hat{y}(\omega)$
&
$\displaystyle \hat{\gamma}(\omega) = J^* (\omega) \, \hat{\sigma}(\omega)$ &
$\displaystyle \hat{Q}(\omega)  = \dfrac{1}{i\omega} Y^* (\omega) \, \hat{V}(\omega) $ &
$\displaystyle \hat{D} (\omega) / \epsilon_0 = \epsilon^*(\omega) \,  \hat{E} (\omega)$ 
\\
\hline 
\end{longtable}
\end{landscape}

\section{Mechanical Spectroscopy}
\label{sec:ms}


Unlike rigid bodies, soft materials deform when a mechanical force is applied and enter a \textit{stressed} state.\cite{shaw2018introduction} The response ranges from (a) an \textit{elastic} solid that remains in the deformed state until the force is removed, thus storing mechanical energy, (b) a \textit{viscous} liquid that deforms continuously while dissipating energy, or (c) a \textit{viscoelastic} response that lies between these two limits. 


In MS, a viscoelastic material is sandwiched between two plates (one stationary and the other mobile) of a rheometer and subjected to a time-dependent strain or stress input. The deflection of the moving plate describes the strain while the force or torque yields the stress. A periodic strain or stress field is applied and the corresponding phase-shifted periodic conjugate variable is measured.\cite{macosko1994rheology, malkin2022rheology} Analysis of the response functions, in either the time or frequency domains, illuminates microstructural relaxation dynamics over different timescales.

Experimental considerations, rather than instrument constraints, dictate the effective frequency range probed. At very low frequencies, a single oscillation cycle requires prohibitively long measurement time, demanding patience, instrument stability, and sample integrity. For instance, at $10^{-6}$ \review{Hz}, a single cycle takes about $10^{6} \text{ s} \approx 11.5 \text{ days}$. This limitation can sometimes be circumvented indirectly using the principle of time-temperature superposition, extending the effective frequency window over many decades.\cite{Dealy2009, Goyal1999} 

\subsection{Mapping with LRT}

In shear rheology, which we use as a proxy for MS, the conjugate variables are shear stress $\sigma \equiv y$ (generalized force) and shear strain $\gamma \equiv x$ (generalized displacement); the qualifier `shear' is dropped hereon. The relaxation modulus $G(t) \equiv R_x(t)$, and the creep compliance $J(t) \equiv R_y(t)$ are material-specific response functions that we seek to learn. The units of $\sigma$ are Pa (pascals) and $\gamma$ is dimensionless. Thus, $G(t)$ and $J(t)$ have units of Pa and Pa$^{-1}$, respectively.

LRT is better known in MS as the Boltzmann superposition principle \addition{(see table \ref{tab:lrt_relations})}.\cite{tschoegl2012phenomenological, cho2016viscoelasticity} It describes the dependence of $\sigma(t)$ on deformation history $\gamma(\tp \leq t)$ or vice versa, where the relationship is modulated by $G(t)$ or $J(t)$, respectively. Owing to the impulse property of response functions, $G(t)$ and $J(t)$ can be obtained from step-strain (stress relaxation) or step-stress (creep) experiments on a system at equilibrium\addition{, respectively}. However, instrument sensitivity  limits the acquisition of reliable $G(t)$ and $J(t)$ signals over very short and long times.\cite{Plazek2000, Shanbhag2024a}

In addition to step inputs, oscillatory shear rheology is a popular technique because the applied perturbation is smooth. The complex modulus $G^{*}(\omega)$ and compliance $J^{*}(\omega)$ are related to the Fourier transforms of $G(t)$ and $J(t)$ via,
\begin{align}
G^{*}(\omega) & = i \omega \hat{G}(\omega)  = \Gp(\omega)  + i \Gpp(\omega) \label{eqn:Gstar_definition}\\
J^{*}(\omega) & = i \omega \hat{J}(\omega) = \Jp(\omega) - i\Jpp(\omega).
\label{eqn:Jstar_definition}
\end{align}
The prefactor $i\omega$ ensures that the complex modulus and compliance retain units consistent with $G(t)$ and $J(t)$, while the sign convention keeps $G^{\prime\prime}$ and $J^{\prime\prime}$ positive for visualization on a log-log plot. This somewhat idiosyncratic convention alters the appearance of familiar LRT relations. For example, eqn. \eqref{eqn:lrt_ft} becomes $\hat{\sigma}(\omega) = G^{*}(\omega) \hat{\gamma}\omg$,  $\hat{\gamma}(\omega) = J^{*}(\omega) \hat{\sigma} \omg$, and $G^{*}(\omega)\, J^{*}(\omega) = 1$.

Note that $G^{\prime}(\omega) \equiv \omega \hat{R}_x^\text{sin}$ and $G^{\prime\prime}(\omega) \equiv \omega \hat{R}_x^\text{cos}$ are called the storage and loss modulus, respectively. Via eqn. \eqref{eqn:oscillatory_lrt}, the stress response to an oscillatory strain ($\gamma(t) = \gamma_0 \sin \omega t$) input is
\begin{equation}
    \sigma(t) = \gamma_0 \left[G^{\prime}(\omega) \sin \omega t + G^{\prime\prime}(\omega) \cos \omega t \right].
\end{equation}
\review{In addition to $G^*$ and $J^*$, a complex viscosity $\eta^* \omg = G^*\omg/i\omega \equiv \hat{R}_x (i\omega)$ is also used.}


\begin{figure}
\begin{center}
\includegraphics[scale=0.6]{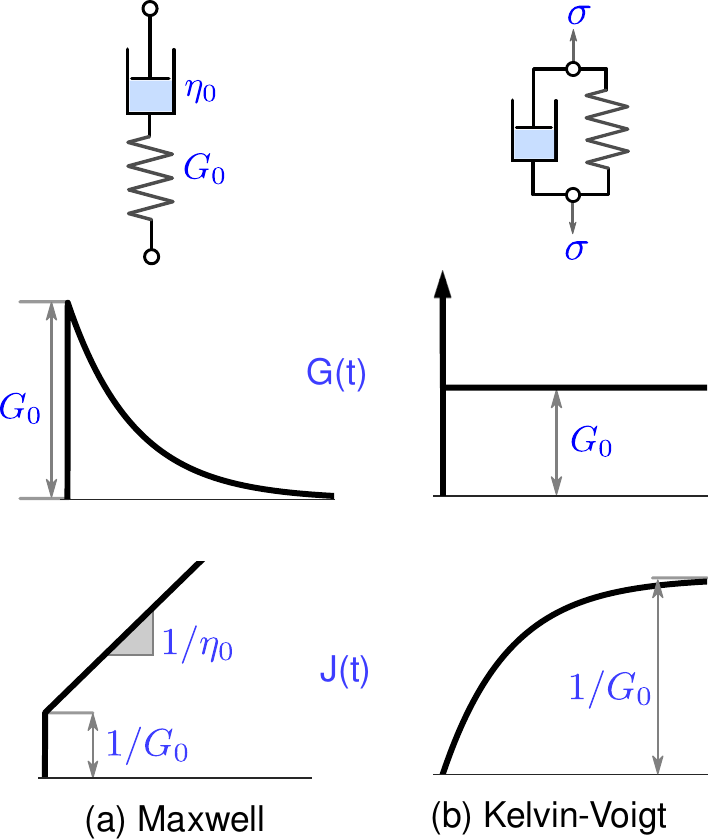}
\end{center}
\caption{Elementary models for MS with a spring and a dashpot in (a) series and (b) parallel. The corresponding response functions are shown schematically in the left and right columns. \label{fig:ms_elementary_models}}
\end{figure}

\subsection{Elementary Models}

The building blocks of models for MS are springs and dashpots. Springs represent elastic and solid-like material attributes responsible for energy storage. On the other hand, dashpots represent viscous and liquid-like attributes responsible for energy dissipation. The stress in a Hookean spring is proportional to strain, $\sigma = G_0 \gamma$, where $G_0$ is the stiffness or modulus. The stress in a dashpot or Newtonian fluid is proportional to strain rate, $\sigma = \eta_0 \dot{\gamma}$, where $\eta_0$ is the viscosity that quantifies resistance to flow. As shown in figure \ref{fig:ms_elementary_models}, these elements can be arranged in series or parallel leading to the Maxwell and Kelvin-Voigt models, respectively.\cite{tschoegl2012phenomenological} The value of these elementary models is not quantitative but qualitative: they illustrate how different arrangements of springs and dashpots capture relaxation phenomena.

\subsubsection{Maxwell Model}

The Maxwell model consists of a spring and a dashpot connected in series (figure \ref{fig:ms_elementary_models}a), ensuring both elements experience the same stress at each instant. The total strain, and hence the total strain rate, is the sum of the elastic and viscous strain rates, $\dot{\gamma} = \dot{\sigma}/G_0 + \sigma/\eta_0$ which can be rearranged into the differential form, 
$\dot{\sigma} + \sigma/\tau = G_0 \dot{\gamma}$, where $\tau = \eta_0/G_0$ is the relaxation time. Using the method of integrating factors, this can be recast into an integral form
\begin{equation}
\sigma(t) = \int_{-\infty}^{t} G_0 e^{-(t-t')/\tau}\,  \dot{\gamma}(t')\, dt'
\label{eqn:maxwell_integral}
\end{equation}
which is a special case of the LRT with $R_x= G(t) = G_0 e^{-t/\tau}$. Mathematical expressions for $G(t)$, $J(t)$, $G^{*}(\omega)$, and $J^{*}(\omega)$ are summarized in Table \ref{tab:ms_properties}.

On application of a step deformation, the spring responds instantly, storing elastic energy. Subsequently, the dashpot starts deforming at the expense of the spring through viscous dissipation (see $G(t)$ in figure \ref{fig:ms_elementary_models}a).\cite{birddpl} Conversely, application of a constant stress results in a sustained, time-dependent deformation, a characteristic of liquid-like behavior (see $J(t)$ in figure \ref{fig:ms_elementary_models}a). The Maxwell model is thus a classic representation for viscoelastic liquids, since $G(t \rightarrow \infty) = 0$ implies complete relaxation. 

In the frequency domain, viscous or terminal relaxation is manifests as $\Gp \sim \omega^2$ and $\Gpp \sim \omega$ for $\omega \tau \ll 1$. Conversely, $\Gp \approx G_0$ and $\Gpp \sim \omega^{-1}$ for $\omega \tau \gg 1$ represents the elastic limit. The crossover of the storage and loss moduli, $\Gp=\Gpp$, which occurs at $\omega \tau=1$, corresponds to the relaxation time.

\begin{table}
\begin{center}
\begin{tabular}{lcccc}
\hline
model & $G(t)$ & $J(t)$ & $G^{*}(\omega)$ & $J^{*}(\omega)$\\
\hline\\[-2ex]
Maxwell & $G_0 e^{-t/\tau}$ &  $\dfrac{1}{G_0} + \dfrac{t}{\eta_0}$ & $\dfrac{G_0 i \omega \tau}{1 + i \omega \tau}$ & $\dfrac{1}{G_0}\left(1 - \dfrac{i}{\omega \tau} \right)$\\[2ex]
Kelvin-Voigt & $G_0 + \eta_0 \delta(t)$ & $\dfrac{1}{G_0} \left( 1 - e^{-t/\tau} \right)$ & $G_0 (1 + i \omega \tau)$ & $\dfrac{G_0^{-1}}{1 + i \omega \tau}$\\[2ex]
\hline
\end{tabular}
\end{center}
\caption{Summary of MS viscoelastic functions for elementary models. \label{tab:ms_properties}}
\end{table}

\subsubsection{Kelvin-Voigt Model}

The Kelvin-Voigt model consists of a spring and a dashpot in parallel so that both elements experience the same strain or deformation at any given time (figure \ref{fig:ms_elementary_models}b). The total stress is the sum of the elastic and viscous contributions, $\sigma = G_0 \gamma + \eta_0 \dot{\gamma}$, which can be cast into a differential form as $\dot{\gamma} + \gamma/\tau = \eta_0^{-1}\sigma$. It may be recast into an integral form, similar to eqn. \eqref{eqn:lrt_reverse0}, with $R_y = J(t) = G_0^{-1} (1 - e^{-t/\tau})$.
\begin{equation}
\gamma(t) = \dfrac{1}{G_0} \int_{-\infty}^{t} \left(1 - e^{-(t-t')/\tau}\right)\,  \dot{\sigma}(t')\, dt'
\label{eqn:kv_integral}
\end{equation}
where $\sigma(t) = \int_{-\infty}^{t} \dot{\sigma}(t')\, dt'$. 

Since the dashpot is arranged in parallel to the spring, the Kelvin-Voigt model is unable to exhibit an instantaneous deformation and $J(t \rightarrow 0) = 0$. Conversely, the spring also restricts the dashpot from deforming infinitely, resulting in a finite steady-state value, $J(t \rightarrow \infty) = G_0^{-1}$. \review{Thus, the Kelvin-Voigt model represents a viscoelastic solid with characteristic \textit{retardation} time $\tau = \eta_0/G_0$ that is conceptually distinct from the \textit{relaxation} time for the Maxwell model.} Interestingly, if the stress is now suddenly removed, the system eventually reverts back to its equilibrium state. This is known as creep recovery and the associated strain is called the recoverable strain.

In the frequency domain, the crossover of $\Jp(\omega)$ and $\Jpp(\omega)$ marks the characteristic retardation time, $\omega \tau = 1$. When stress (force) is applied to a viscoelastic solid, the dashpot resists rapid deformation, thereby delaying or retarding the system’s progression to a new equilibrium state. As a result, viscous dissipation dominates the short-time response ($\omega \tau \gg 1$, $\Jp < \Jpp$), while elastic, spring-like behavior governs the long-time response ($\omega \tau \ll 1$, $\Jp > \Jpp$).

\section{Electrochemical Impedance Spectroscopy}
\label{sec:eis}

EIS is an analytical technique that provides mechanistic insights into electrochemical systems by deconvoluting processes occurring at different timescales such as rapid charge transport, intermediate double-layer effects, and slow diffusion processes. In an experiment, a system is perturbed using a small alternating voltage  (potentiostatic) or current (galvanostatic) \addition{with an optional offset or DC bias}, and the resulting phase-shifted sinusoidal current or potential is measured over a range of frequencies. \review{The DC bias sets the electrochemical operating point on the current-potential curve, while the alternating perturbation probes the impedance response at that condition.}

\review{A typical electrochemical cell uses a three-electrode configuration -- working, counter, and reference. A potentiostat adjusts the potential between working and reference electrodes to control the reaction, while current flows between the working and counter electrodes. Alternately, different electrodes may be used to study specific ions or chemistry. Modern EIS setups employ a potentiostat–galvanostat with a frequency response analyzer that measures the amplitude ratio and phase shift between the current and voltage at each frequency.}
A transfer function called electrical impedance is defined as the ratio of the Fourier transforms of the voltage and current,\cite{wang2021electrochemical}  
\begin{equation}
Z^{*}(\omega) \equiv \dfrac{\hat{V}(\omega)}{\hat{I}(\omega)}.
\label{eqn:impedance_defn}
\end{equation}
The units of $V$, $I$, and $Z^{*}$ are volts, amperes, and ohms, respectively. Impedance generalizes the concept of resistance to time-varying systems. For consistency with MS and BDS, we retain the superscript `*' to denote frequency domain properties although it is frequently dropped. Once the impedance spectrum $Z^*\omg$ is obtained, it is typically analyzed by fitting an equivalent circuit composed of resistors $(R)$, capacitors $(C)$, inductors $(L)$, and other specialized elements such as Warburg impedance and constant phase element that correspond to specific electrochemical phenomena.


The usable frequency range in an EIS experiment is regulated by the stability of the electrochemical system (low frequencies), instrument wiring (high frequencies), and the timescales of electrochemical processes anticipated in the sample. As in MS, measurements at very low frequencies may take hours or even days per data point, whereas a standard EIS frequency sweep typically requires only a few minutes.


\subsection{Mapping with LRT}

While EIS is analyzed using voltage and current, the conjugate variables that map to the LRT are voltage (generalized force) and charge $Q = \int I(t) dt$ (generalized displacement). Thus, EIS works with $\dot{x} = \dot{Q} = I$ in practice. From eqn \eqref{eqn:impedance_defn} and the first relation in eqn \eqref{eqn:lrt_ft}, the impedance $Z^{*}(\omega) = \hat{V}/\hat{\dot{Q}} \equiv \hat{R}_x(\omega)$. Since $\Zstar\omg$ is not the ratio of force and displacement, it cannot be thought of as a generalized modulus. Its counterpart, the admittance $Y^*\omg=1/Z^*\omg=-\omega^2 \hat{R}_y\omg$, due to the convolution relation.  To summarize, the LRT relations (eqn. \eqref{eqn:lrt_ft}) for EIS in the Fourier domain are, 
\begin{align}
    \hat{V}\omg &=i \omega \, \Zstar\omg \hat{Q}\omg = \Zstar\omg \hat{I}\omg, \\
    \hat{I}\omg &= i \omega \hat{Q}\omg = Y^* \omg \hat{V}\omg .
\end{align}
The time domain response functions $R_x(t)$ and $R_y(t)$ follow from inverse Fourier transforms but are seldom used in impedance spectroscopies, which emphasize frequency domain analysis.

Consider galvanostatic EIS where $I(t)=I_0 \cos{\omega t}$ is applied, and the potential $V(t)=V_0 \, \cos{(\omega t+ \delta)}$  phase shifted by $\delta(\omega)$ is measured.  Recasting eqn. \eqref{eqn:impedance_defn} into modulus and argument format,
\begin{equation}
Z^{*}(\omega) = \left| \dfrac{\hat{V}(\omega)}{\hat{I}(\omega)}\right|\left(\cos{\delta + i\sin{\delta}} \right)= Z'(\omega) + i Z''(\omega).
\label{eqn:impedance_defn2}
\end{equation}
The component $\Zp \omg= (V_0/I_0) \cos{\delta}$ (resistance) is in-phase with current and represents energy dissipated as heat, while $\Zpp \omg= (V_0/I_0) \sin{\delta}$ (reactance) is out-of-phase and reflects temporary energy storage in electric (capacitive) or magnetic (inductive) fields. \textit{This is an important distinction}: $\Zp$ and $\Zpp$ denote energy loss and storage in EIS, \review{which is a reversal of roles played by ${}^\prime$ and ${}^{\prime \prime}$ in MS.} Furthermore, the complex admittance $Y^*$ reflects the same information as $Z^*$, and is rarely used unless  very small or large impedances are encountered.

EIS measurements are usually analyzed in terms of:
\begin{enumerate}[label=(\roman*)]
\item Nyquist plots of $-\Zpp\omg$ versus $\Zp\omg$ on the Argand plane
\item Bode plots of magnitude  $|\Zstar\omg|=\sqrt{Z^{\prime \, 2} +Z^{\prime \prime \, 2}}$ versus $\omega$
\item Bode plots of phase angle $-\delta \omg$ versus $\omega$.
\end{enumerate}
Distinct features of these plots are associated with specific phenomena and represented with appropriate circuit element(s).\cite{Lasia2014} The goal is to assemble these elements into an \textit{equivalent circuit} to model the entire electrochemical system.

\subsection{Elementary Models}

The building blocks of electric circuits are resistors (resistance $R$), capacitors (capacitance $C$), and inductors (inductance $L$). The units of $R$, $C$, and $L$ are ohm $(\Omega)$, farad $(F)$, and henry $(H)$, respectively.
\begin{enumerate}[label=(\roman*)]
    \item Resistors are dissipative elements that allow a continuous flow of charge (current) on application of a voltage and do not store electrical energy. For resistors $V =  R \dot{Q}$ (Ohm's law), and $Z_R^{*} = R = \Zp_R$ is real and independent of frequency.

    \item Capacitors are passive elements that store electrical energy as an electric field between two conductive plates separated by an insulator or dielectric material. For capacitors $V = Q/C$ and $Z_C^{*}(\omega) = 1/(i \omega C)$ is purely reactive with $\Zp_C = 0$ and $\Zpp_C = -1/(\omega C) < 0$. The current leads voltage by $90^\circ$ and $|Z_C^{*}|$ decreases as $\omega$ increases. As $\omega \rightarrow 0$, the $V \rightarrow V_0$ is constant, which causes the capacitor to behave like a pure insulator and drives impedance $|Z_C^{*}| \rightarrow \infty$.

    \item Inductors store energy in the form of a magnetic field due to the flow of changing current through a coil. They act as inertial elements with  $V = L \dot{I} = L \ddot{Q}$ and  $Z_L^{*}(\omega) = i \omega L$. The current lags the voltage by $90^\circ$. Unlike a capacitor, $|Z_L^{*}|$ increases with $\omega$.
\end{enumerate}

These basic blocks represent ideal behavior that is rarely observed in real materials. For instance, even a copper wire, which is often considered a pure resistor, shows frequency-dependent impedance due to eddy currents. Nonetheless, such elements serve as conceptual proxies for real electrochemical processes. For example, a resistor is a useful approximation for solution resistance and charge transfer resistance across the electrode-electrolyte interface. Similarly, the electrical double layer at electrode surfaces inherently behaves like a capacitor.\cite{Lasia2002,Lasia2014} Self-inductance usually appears at high frequencies because rapidly changing currents generate magnetic fields. In contrast, impedance spectra show low-frequency inductive loops that arise from the adsorption of reaction intermediates.\cite{Lazanas2023,wang2021electrochemical}


\begin{figure}
\begin{center}
\includegraphics[scale=1.7]{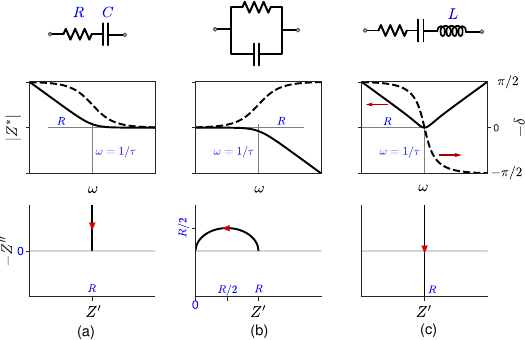}
\end{center}
\caption{The top row shows (a) RC series, (b) RC parallel, and (c) RLC circuits. The second and third rows show the Bode and Nyquist plots (linear plots of $-\Zpp$ versus $\Zp$), respectively. The Bode plots of magnitude $|Z^{*}|$ versus $\omega$ (solid lines) use a double logarithmic scale, while the Bode plots of phase angle (dashed lines) use a linear scale for phase angle. Important landmarks are indicated on the plots. In the Nyquist plots, the arrow points in the direction of increasing $\omega$. Plots in a given row use a shared vertical axis. \label{fig:eis_elements}}
\end{figure}

\subsubsection{Series and Parallel Circuits}

We now examine the impedance spectra of equivalent circuits formed by connecting these basic elements. Since impedance is a generalized resistance, the impedance of elements connected in series is additive. When elements are connected in parallel, the admittance ($1/Z^*$) is additive. 

\begin{enumerate}[label=(\roman*)]
\item \textit{RC Series}: When a resistor and capacitor are connected in series (figure \ref{fig:eis_elements}a) the effective impedance is
\begin{equation}
Z^{*}_\text{RC, series} = Z_R^{*} + Z_C^{*} = R - i\dfrac{1}{\omega C}.
\end{equation}
The characteristic timescale $\tau = RC$ separates the high frequency $\omega \tau \gg 1$ behavior, where the response is predominantly resistive ($|Z^{*}|/R \approx 1$ and $\delta \to 0$ in figure \ref{fig:eis_elements}a) from the low frequency $\omega \tau \ll 1$ behavior, where the response is predominantly capacitive ($|Z^{*}|/R \approx (\omega \tau)^{-1} $ and $\delta \to -90^\circ$).  The Nyquist plot (last row figure \ref{fig:eis_elements}a) appears as a vertical line at $\Zp = R$ on the real axis. 

\item  \textit{RC parallel}: When a resistor and capacitor are connected in parallel (figure \ref{fig:eis_elements}b) the effective impedance is
\begin{equation}
Z^{*}_\text{RC, parallel} = \left( \dfrac{1}{Z_R^{*}} + \dfrac{1}{Z_C^{*}} \right)^{-1} = \dfrac{R}{1 + i \omega \tau} = \dfrac{R}{1 + \omega^2 \tau^2} - i \dfrac{R \omega \tau}{1 + \omega^2 \tau^2},
\end{equation} 
where $\tau=RC$ is the relaxation time, the timescale over which the circuit charges or discharges. On the Bode plots (figure \ref{fig:eis_elements}b), $|Z^*|/R \approx  (\omega \tau)^{-1}$ when $\omega \tau \gg 1$ indicating reduced opposition to current flow. At low frequencies, the impedance is predominantly resistive $|Z^*| \approx R$ as the current avoids the capacitor which behaves like an insulator. The Nyquist plot (figure \ref{fig:eis_elements}b) is a semicircle located in the upper half of the complex plane, where $(\Zp, \Zpp) = (0, 0)$ corresponds to $\omega \tau \to \infty$ and $(\Zp, \Zpp) = (R, 0)$ to $\omega \tau \to 0$. 
Overlapping semicircles in a Nyquist plot are attributed to distinct relaxation processes.

\item \textit{RLC series}: The effective impedance of the series RLC element is
\begin{equation}
Z^{*}_\text{RLC, series} = Z_R^{*} + Z_C^{*} + Z_L^{*} = R + i\left(\omega L - \dfrac{1}{\omega C}\right).
\end{equation} 
At the resonance frequency $\omega_\text{res} = 1/\sqrt{LC}$, the reactive component of $\Zstar$ is zero; the impedance is minimized, which allows maximum current flow. This minimum in $|\Zstar|$ is also evident in the Bode plot (figure \ref{fig:eis_elements}c). 
The system is predominantly capacitive ($\Zpp < 0, \delta<0$) for $\omega \tau \ll 1$ and equivalent to a series RL circuit ($\Zpp > 0,\delta>0$) for $\omega \tau \gg 1$. The Nyquist plot is a straight line at $\Zp=R$, extending from $-\Zpp \to \infty$ at small $\omega$ (capacitive) to $-\Zpp \to -\infty$ at large $\omega$ (inductive).

The relative dearth of inductive elements to capacitive elements in equivalent circuits explains why $-\Zpp$ and $-\delta$ are used as a convention in the Bode and Nyquist plots instead of $\Zpp$ and $\delta$.

\end{enumerate}

\subsubsection{Other Models}

Besides these building blocks, EIS uses specialized circuit elements to mechanistically model electrochemical processes. For example, a Warburg element, with impedance $Z^{*} = \sigma_{W} (\omega^{-1/2} - i \omega^{-1/2})$, may be combined with other basic elements to form a Randles circuit, which describes Faradic processes in a redox reaction. The Warburg element can be visualized as a frequency-dependent series RC circuit. The constant phase element is another flexible element (analogous to a springpot in MS) with impedance $\Zstar \sim (i\omega)^{-n}$ with $0 \leq n \leq 1$. It behaves like a resistor when $n=0$, a Warburg element when $n=0.5$, and a capacitor when $n = 1$. 

Equivalent circuits are non-unique; the same data may be described by different numbers, permutations, and types of circuit elements. \review{Non-uniqueness extends to the fitted parameters, which are estimated via nonlinear least-squares; it is further exacerbated by stochastic and bias errors in the experimental data.\cite{orazem2020tutorial, ZHANG2015464, Lasia2014, orazem2008electrochemical}}
Since each element maps to a distinct process, identifying an appropriate equivalent circuit is more important in EIS than in MS or BDS. 

\section{Broadband Dielectric Spectroscopy}
\label{sec:bds}

Consider \review{an ideal} capacitor in which a dielectric material is sandwiched between metallic parallel plates with surface area $A$ separated by a distance $d$. When a constant potential difference $V$ is applied across the capacitor, no current flows \review{(in principle)} due to the bulk motion of charge carriers\review{, unlike a resistor}. Instead, the capacitor stores energy by creating a separation of charges across the plates. This electrical charge $Q = CV$, where the capacitance $C = \epsilon C_0$. $\epsilon$ is the relative permittivity, $C_0  = \epsilon_0 A/d$ is the capacitance of vacuum, and $\epsilon_0 = 8.854 \times 10^{-12}$ F/m is the permittivity of free space.

\rereview{Dipoles in a dielectric material respond to the imposed electric field $E = V/d$ (V/m). The displacement of electron clouds and the relative motion of oppositely charged ions occur over fast timescales ($\lesssim 10^{-12}$ s) and collectively contribute to \textit{induced polarization}. Molecular dipoles, on the other hand, reorient on a much slower timescale leading to \textit{orientational polarization}. This is the primary relaxation mechanism accessible in BDS experiments. In addition, in heterogeneous systems, the accumulation of free charges at internal boundaries results in interfacial polarization.} 
From Gauss' law, the dielectric displacement $D = Q/A$ (Cm$^\text{-2}$) is equal to the surface charge density. When $E$ is small, the response is linear and $D = \epsilon \epsilon_0 E$\rereview{, implying a displacement current density $j_d=\dot{D}=\epsilon \epsilon_0 \dot{E}$. However, in practice, unlike an ideal dielectric, additional conduction currents arise due to bulk motion of mobile charges with a current density $j_c = \sigma_\text{dc} E$. Thus, the total current density $j=j_c+j_d=\sigma_\text{dc} E+\epsilon \epsilon_0 \dot{E} $. }

\review{The dielectric sample is sandwiched between two flat electrodes like a capacitor. Unlike an EIS setup, the system does not use a reference electrode or employ a DC bias.} When a potential difference $V(t) = V_0 \sin \omega t$ is applied, it imposes a periodic electric field $E(t) = E_0 \sin \omega t$ that elicits a periodic (phase-shifted) $D(t)$ resulting in a sinusoidal current. 
Using the applied voltage and measured current, modern commercial dielectric spectrometers yield the real and imaginary parts of the impedance $Z^{*}(\omega)$. The primary goal of BDS experiments is to characterize the complex permittivity,
\begin{equation}
\epsilon^{*}(\omega) = \dfrac{1}{i \omega C_0 Z^{*}(\omega)} =\epsp\omg-i \epspp\omg.
\label{eqn:epsilon_star_impedance}
\end{equation}
where, $\epsp\omg$, $\epspp\omg$, and $\tan{\delta}=\epspp/\epsp$ are the dielectric storage, dielectric loss, and loss tangent, respectively. 

BDS probes a broad range of frequencies below the infrared regime, from the $\mu$Hz range up to optical frequencies. It encompasses several complementary subfields, each addressing the dielectric response within a portion of this spectrum. Collectively, these approaches provide the extensive frequency coverage that explains the label `broadband' in BDS.\cite{woodward2021broadband, kremer2002broadband}


\subsection{Mapping with LRT}

The conjugate variables of BDS are $y(t) \equiv E(t)$ (generalized force) and $x(t) \equiv D(t)/\epsilon_0$ (generalized displacement).  Interestingly, both these variables have the same units (V/m). 
The key response function in BDS is the dielectric permittivity $\epsilon(t) \equiv R_y(t)$ which can be thought of as a generalized compliance. It is useful to partition $\epsilon(t)$ into two parts,
\begin{equation}
\epsilon(t) = \epsilon_{\infty} + \Delta \epsilon \left(1 - \phi(t)\right),
\label{eqn:epsilont}
\end{equation}
where the first term corresponds to the ``instantaneous'' induced polarization, where subscript ``$\infty$'' refers to the high-frequency (fast) response by convention, and $\epsilon_\infty=\epsilon(t=0)=\epsp(\omega \to \infty)$. The second term captures the slower dipolar rearrangement due to molecular reconfiguration and $\Delta \epsilon = \epsilon_s - \epsilon_{\infty}$, where $\epsilon_s = \epsilon(t \to \infty) = \epsp(0)$ is the static or low frequency permittivity (the subscript `s' is used instead of `0' to avoid confusion with the permittivity of vacuum). The dimensionless monotonic relaxation function,
\begin{equation}
\phi(t) = \dfrac{\epsilon_s - \epsilon(t)}{\epsilon_s - \epsilon_{\infty}}
\label{eqn:phi_dielectric}
\end{equation}
decreases from $\phi(0)=1$ to $\phi(t \to \infty)=0$.  $\epsilon(t)$, $\phi(t)$, $\epsilon_{\infty}$, $\epsilon_{s}$, and $\Delta \epsilon$ are all dimensionless. Analogous to $J^*\omg$ in MS, the complex permittivity is given by,
\begin{equation}
    \epsilon^{*}(\omega) \equiv \dfrac{\hat{D}(\omega)}{\epsilon_0 \hat{E}(\omega)} = i \omega \hat{\epsilon}(\omega) = \epsilon(0)+\hat{\dot{\epsilon}} = \epsilon_\infty-\Delta\epsilon \,  \hat{\dot{\phi}}, \label{eqn:estar_from_et}
\end{equation}
where the last relation follows from eqn \eqref{eqn:epsilont}. Thus, $\epsilon^{*}(\omega)$ can be decomposed into real and imaginary parts,\cite{lee2010multiple}
\begin{align}
\epsilon^{\prime}(\omega) & 
=\epsilon_{\infty}  - \Delta \epsilon \int_{0}^{\infty} \dot{\phi}(s)\,  \cos (\omega s) \, ds \notag\\
\epsilon^{\prime\prime}(\omega) & = - \Delta \epsilon \int_{0}^{\infty} \dot{\phi}(s)\,  \sin (\omega s) \, ds.
\end{align}
The complex dielectric modulus $M^{*}(\omega) = 1/\epsilon^{*}(\omega)$ contains equivalent information but is used less frequently. For studies of slow dielectric relaxation processes due to motion of molecules and ions in homogeneous materials $\epsilon^{*}$ is generally considered to be superior. 

\begin{figure}
\begin{center}
\includegraphics[width=0.45\textwidth]{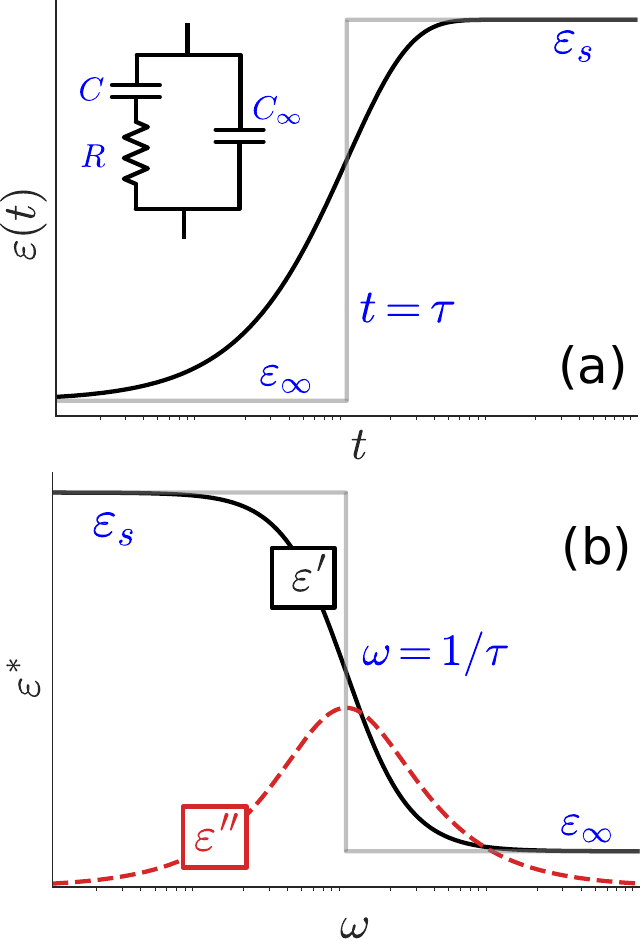}
\end{center}
\caption{The (a) time domain $\epsilon(t)$ and (b) frequency domain $\epsstar$ permittivity for the Debye model. The inset in (a) shows the equivalent circuit where the capacitor $C_\infty$ models induced polarization. The horizontal axes are logarithmic.} 
\label{fig:bds_elements}
\end{figure}

\subsection{Elementary Models}
\label{sec:bds_elementary_models}

BDS uses capacitors and resistors as building blocks that model energy storage and dissipation due to leakage, respectively. A capacitor represents an ideal insulator that gets polarized and exhibits a dielectric displacement. In contrast, a resistor models the bulk motion of charge carriers that lead to sustained conduction currents. Unlike EIS, inductors are generally not used in BDS because the small currents involved (\review{on the order of  pA-$\mu$A}) do not generate appreciable magnetic energy storage.\cite{kremer2002broadband} The impedance of  capacitors $Z_C^*(\omega) = (i \omega C)^{-1}$ and resistors $Z_R^*(\omega) = R$ leads to $\epsstar_C=\epsilon$ and $\epsilon_{R}^{*} = (i \omega R C_0)^{-1}$ via eqn \eqref{eqn:epsilon_star_impedance}, respectively.

The frequency dependence of dielectric permittivity stems from distinct mechanisms operating on different timescales, namely rotational diffusion in molecular dipoles, translational diffusion of mobile charges, separation of charges at interfaces, etc. These may be modeled as equivalent circuits of resistors and capacitors. For example, the series RC circuit discussed previously (figure \ref{fig:eis_elements}a) accounts for dipolar relaxation with permittivity,
\begin{equation}
    \epsstar \omg = \dfrac{\epsilon}{1+i \omega \tau}, \label{eqn:estar_dipole}
\end{equation}
where $\tau = RC=\epsilon C_0 R$. This is analogous to $J^{*}$ for a Kelvin-Voigt element in MS (Table \ref{tab:ms_properties}). 
Equivalent circuits are useful for visualizing simple materials with nearly Debye-type relaxation or for instrument-related corrections (stray capacitance, resistance, or inductance).\cite{woodward2021broadband} However, they are not fundamental representations of dielectric relaxation in complex systems, where empirical models are commonly used.

\subsubsection{Debye Model}

The Debye model considers the relaxation of non-interacting dipoles. It assumes a first order process with relaxation function $\phi(t) = e^{-t/\tau}$. This implies a permittivity, $\epsilon(t) =\epsilon_\infty+\Delta\epsilon \left( 1 - e^{-t/\tau} \right)$ and a complex permittivity, 
\begin{equation}
\epsilon_{\text{Debye}}^{*}(\omega) = \epsilon_{\infty}  + \dfrac{\Delta \epsilon}{1 + i \omega \tau}.
\label{eqn:debye_estar}
\end{equation}
Separating the real and imaginary parts yields,
\begin{align}
\epsilon_{\text{Debye}}^{\prime}(\omega) & = \epsilon_{\infty}  + \dfrac{\Delta \epsilon}{1 + \omega^2 \tau^2}, \label{eqn:debye_epsp} \\
\epsilon_{\text{Debye}}^{\prime\prime}(\omega) & =  \dfrac{\Delta \epsilon \, \omega \tau}{1 + \omega^2 \tau^2}.  \label{eqn:debye_epspp}
\end{align}
Figures \ref{fig:bds_elements}a and \ref{fig:bds_elements}b show $\epsilon(t)$ and $\epsstar\omg$ for the Debye model, respectively. 
With increasing frequency $\epsp\omg$ shows a sigmoid-like step decrease, while $\epspp\omg$ shows a peak at $\omega \tau = 1$. This loss peak is symmetric with a half-width of about 1.14 decades. The dielectric strength $\Delta \epsilon$ can be found either from the step size in $\epsp$ or the area under the $\epspp$ curve. The equivalent circuit that corresponds to the Debye model is shown in figure \ref{fig:bds_elements}a. Its impedance is,
\review{\begin{equation}
    \Zstar_\text{Debye} =  \left[ \left(R + (i \omega C)^{-1} \right)^{-1} + i \omega C_\infty \right]^{-1}, \nonumber
\end{equation} }
where $\epsilon_\infty = C_\infty/ C_0$, $\Delta \epsilon = C/C_0$, $\epsilon_s=(C+C_\infty)/C_0$, and $\tau=RC$. $C_\infty$ and $C_s =  C + C_\infty$ are capacitances that correspond to induced polarization and a fully charged system, respectively. 


\subsubsection{Havriliak-Negami}
\label{sec:bds_HNm}

\begin{figure}
\begin{center}
\includegraphics[scale=0.6]{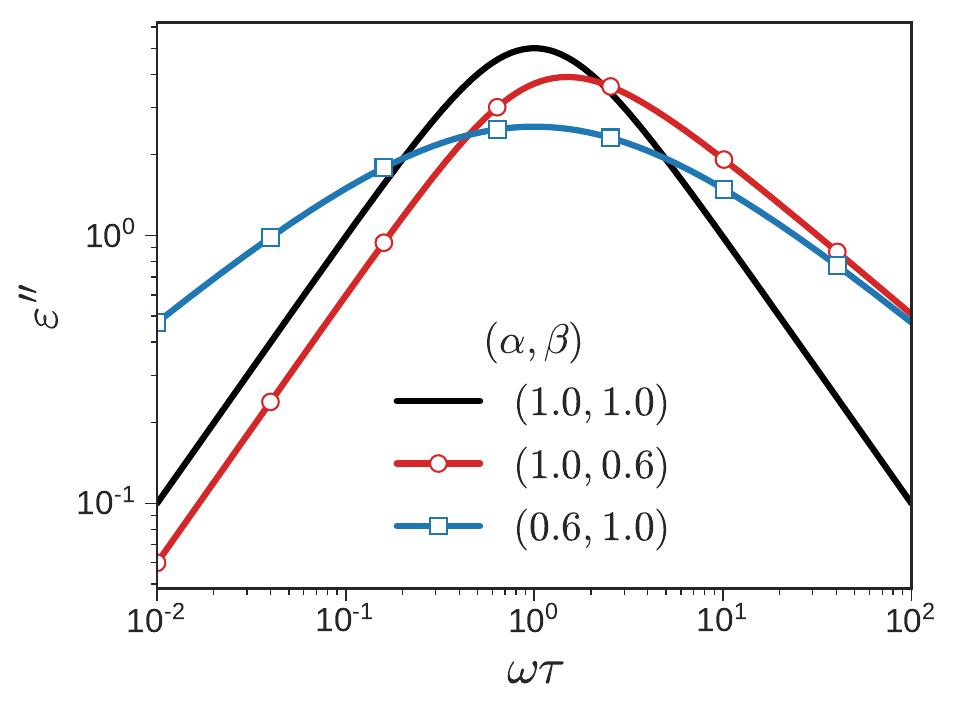}
\end{center}
\caption{$\epspp\omg$ for the Havriliak-Negami function at different $\alpha$ and $\beta$. The Debye model is a special case with $\alpha=\beta=1$. $\alpha$ and $\beta$ control the broadening around the peak and near the high frequency tail, respectively. }
\label{fig:estar_HN}
\end{figure}

Although the Debye model captures the essential features of dielectric relaxation, real materials often show broader, asymmetric loss peaks (on plots of $\epspp$ versus $\omega$) than predicted by exponential relaxation. They are typically fitted to the Havriliak-Negami (HN) equation, an empirical generalization of the Debye model,
\begin{equation}
\epsilon_{\text{HN}}^{*}(\omega) = \epsilon_{\infty}  + \dfrac{\Delta \epsilon}{\left(1 + (i \omega \tau)^{\alpha}\right)^{\beta}},
\label{eqn:hne_estar}
\end{equation}
with $0 < \alpha, \beta \leq 1$. It accounts for the frequently observed power-law dependence of $\epsp$ and $\epspp$ on $\omega$.

Figure \ref{fig:estar_HN} shows a comparison of the $\epspp\omg$ for the HN function at different values of $\alpha$ and $\beta$ with the Debye model. Decreasing $\alpha$ symmetrically broadens $\epspp\omg$ around its peak. Decreasing $\beta$ asymmetrically broadens the high-frequency tail. Thus, the HN equation provides a convenient way to fit experimental data over a broad frequency range instead of a large number of Debye modes. Depending on the number of peaks observed in $\epsilon^{\prime\prime}$, more than one HN mode may be justified. 

\subsubsection{Conduction Currents}


Besides dipolar relaxation, conduction currents arise from the motion of mobile charge carriers under an applied field. The parallel RC (figure \ref{fig:eis_elements}b) circuit is representative of pure electronic conduction, with the resistor $(R=d/A\sigma_\text{dc})$ capturing dissipative effects of migrating charges while the capacitor $(C=\epsilon_0\epsilon_\infty A/d)$ accounts for induced polarization. This yields, 
\begin{equation}
   \epsstar \omg=\epsilon_\infty - i \dfrac{\sigma_\text{dc}}{\epsilon_0 \omega}. \label{eqn:bds_conduction}
\end{equation}
\review{This relation assumes that charges move freely, independently, and without memory. However, for most real disordered systems like ionic conductors, glasses, and polymers, charge hopping is an activated process that violates these assumptions. Instead of $\epsilon^{\prime\prime}(\omega) \sim \omega^{-1}$, these materials exhibit $\epsilon^{\prime\prime}(\omega) \sim \omega^{-n}$, with $0 < n < 1$, which is called universal dielectric response.\cite{Jonscher1977, Dyre2000} In any event, strong conduction currents result in large $\epspp$ at small $\omega$ that mask the signal from orientational polarization. It may be advisable to analyze such data using $M^*\omg$ to isolate this effect since it manifests as a separate peak in $M^{\prime \prime}\omg$.\cite{tian2014electric, ohki2022broadband}}


\section{Discussion}
\label{sec:discussion}

Despite their origins in the disparate disciplines of rheology, electrochemistry, and dielectric physics, all three spectroscopic techniques ultimately decompose system responses into storage and dissipation of energy. Such a unified view not only aids pedagogy, but also enables cross-pollination of ideas and methods. MS, EIS, and BDS probe systems with oscillatory stimuli to characterize LRT response functions in the time or frequency domain.  EIS and BDS are impedance spectroscopies that aim to quantify circuit impedance using alternating voltages or currents.
\rereview{They share similar instrumentation and theoretical foundations but serve different roles: BDS analyzes the molecular mechanisms that control orientation polarization and charge transport in dielectrics, while EIS investigates kinetics in electrochemical systems through lumped circuit representations. }
\review{Both use a sample between electrodes, a signal generator, and a frequency response analyzer, and can sometimes substitute for each other, albeit within limited frequency and temperature ranges.} 
On the other hand, MS uses distinct terminology and instrumentation. Consequently, the subsequent discussion centers around MS and impedance spectroscopy (IS); distinctions between EIS and BDS are highlighted wherever appropriate.

\subsection{Building Blocks}
\label{Sec:elements_discussion}


\renewcommand{\arraystretch}{2} 
\begin{table}
    \centering
    \begin{tabular}{l|cc}
    \hline
        \textbf{Attribute} & \textbf{MS} & \textbf{IS}
        \\
        \hline\\[-5ex]        
         \multirow{2}{*}{storage element} & 
        \includegraphics[height=0.03\textwidth]{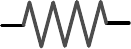} &
        \includegraphics[height=0.04\textwidth]{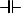}  
        \\[-2ex]
        & $\sigma= G \gamma$  & $V=C^{-1} Q$  \\
         \multirow{2}{*}{dissipation element} & 
		 \includegraphics[height=0.03\textwidth]{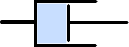} &
         \includegraphics[height=0.03\textwidth]{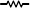} 
         \\ [-2ex]
         & $\sigma =\eta \dot{\gamma}$ & $V = R\dot{Q}$  \\
\hline
Stiffness $(k)$ & $G$ & $C^{-1}$  \\
Viscosity $(\beta)$ & $\eta$ & $R$  \\
$\tau$ $(\beta/k)$   & $\eta/G$ & $RC$  \\
\hline 
    \end{tabular}
    \caption{Energy storage and dissipation units in MS and IS.}
    \label{tab:fundamental_elements}
\end{table}

The building blocks of elementary models in MS and IS are energy  \textit{storage} and \textit{dissipation} elements, which represent two idealized limits for how nature partitions energy (see Table \ref{tab:fundamental_elements}). Storage elements capture the instantaneous response to an excitation, while the dissipation elements account for delayed, time-dependent behavior.

\begin{enumerate}
    \item \textbf{Energy Storage Elements}: These units, represented by springs (MS) and capacitors (IS), store external work as potential energy through a change in state, which may be fully recovered upon unloading.  The restoring force $y$ is proportional to the displacement $x$, $y = kx$. The response is instantaneous and in-phase with the input; its magnitude is governed by the stiffness $k$. 
    In MS, this is identified with ideal elastic or rubbery behavior given by $\sigma = G\gamma$, where $\sigma \equiv y$, $G \equiv k$, and $\gamma \equiv x$ (the subscript `0' in $G_0$ and $\eta_0$ is dropped hereon for brevity). In IS or electrical circuits, this role is performed by ideal capacitors which follow $V=Q/C$, where $V \equiv y$, $1/C \equiv k$, and $Q \equiv x$. The restoring force generated in these elements is conservative \review{and the energy stored $k x^2/2$ is path-independent. Consequently, for cyclic oscillations, the net energy storage -- given by the in-phase component -- is zero, as the system returns to its original state after each cycle; see eqn. \eqref{eqn:energy_storage}. }
    

    \item \textbf{Energy Dissipation Elements}: These units dissipate  external work via irreversible mechanisms. This leads to continuous displacement under the action of a force, $y= \beta  \dot{x}$, where $\beta$ quantifies resistance to flow; $y$ is out-of-phase with $x$ (it is in phase with velocity $\dot{x}$). In MS, this is identified with a dashpot or Newtonian fluid $(\sigma = \eta \dot{\gamma})$ that dissipates energy as heat via internal friction between fluid layers.\cite{bird2002transport} The analogous element in IS is a resistor with $V = R\dot{Q}$, where friction between flowing charges and the conductor produces thermal dissipation called Joule heating.\cite{Shankar2020} $\eta$ and $R$ are equivalent to $\beta$ in MS and IS, respectively. Since forces generated in these elements are non-conservative, work done on the system is irreversibly lost to the environment. Unlike storage elements, the work done in dissipation units $W = \int \beta \dot{x} \, dx = \int \beta (\dot{x})^2 \, dt$ is path-dependent, and hence non-zero over a closed cycle; \review{see eqn. \eqref{eqn:energy_dissipate}}. 
    
\end{enumerate}

\review{Interpreting BDS responses in terms of resistors and capacitors is less straightforward than EIS, since the conjugate variables are electric field and displacement rather than voltage and current. Conceptually, a capacitor maps to an ideal insulator that stores energy via polarization, while a resistor maps to a conductor that permits charge movement. By analogy with MS, the spring and dashpot correspond to the insulator and conductor, respectively. This framing can be found in the literature, although it is not standard in conventional BDS analysis.\cite{kremer2002broadband, shaw2018introduction}}  Furthermore, in EIS, we sometimes encounter inductive elements which oppose changes in current, $V = L \dot{I} = L\ddot{Q}$. They behave like  inertial elements $y = m \ddot{x}$, where $m$ is mass. MS and BDS techniques aim to avoid inertial effects, and hence we do not discuss them further. 


\subsection{Combinations of Building Blocks}

\begin{table}
    \centering
    \begin{tabular}{cc}
    \hline
         \textbf{MS} & \textbf{IS}
         \\ \hline \\ [-4ex] 
         
         \raisebox{-1.5ex}{\includegraphics[height=0.03\textwidth]{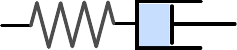} } 
         & \raisebox{-3ex}{\includegraphics[height=0.07\textwidth]{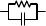}} \\
        
         Maxwell model  
         & Parallel RC 
         \\ [2ex]
 \raisebox{-3ex}{\includegraphics[height=0.07\textwidth]{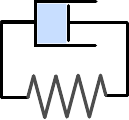}} 
         & \raisebox{-2ex}{\includegraphics[height=0.04\textwidth]{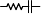}} 
         \\
         Kelvin-Voigt model
         & Series RC
         \\
         \raisebox{-5ex}{\includegraphics[height=0.08\textwidth]{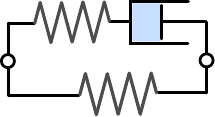}}
         & \raisebox{-5ex}{\includegraphics[height=0.08\textwidth]{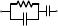}} 
         \\ [2ex]
 Zener model  & \\ 
		\raisebox{-5ex}{\includegraphics[height=0.08\textwidth]{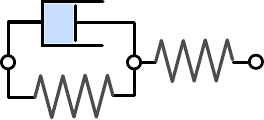}}  
         &  \raisebox{-5ex}{\includegraphics[height=0.09\textwidth]{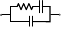}}
         \\ 
  Poynting-Thompson model & Debye model \\
         \hline
    \end{tabular}
    \caption{Equivalent circuits in MS and IS.}
    \label{tab:combination_elements}
\end{table}

The mathematical equivalence of building blocks for energy storage and dissipation in MS and IS does not automatically propagate to their series and parallel combinations (see Table \ref{tab:combination_elements}). It is helpful to compare the same material function, say $\hat{y}(\omega)/i\omega \hat{x}(\omega)= \hat{R}_x$ (eqn. \eqref{eqn:lrt_ft}), for such mechanical and electrical systems. In MS, $\hat{R}_x= G^{*} \omg/i\omega = \eta^*\omg$, while in IS $\hat{R}_x = Z^{*}(\omega)$. Thus, $\eta^{*}$ and $Z^{*}$ are analogous ($\cong$), i.e.,
\begin{equation}
\dfrac{G^{*}}{i\omega} = \eta^* \cong  Z^{*}.
\end{equation}
Since all linear material functions can be obtained from one another (eqn. \eqref{eqn:lrt_ft}), the equivalence of all analogous relations follows (section \ref{sec:response}). 

The response of a series spring and dashpot (Maxwell model) resembles that of a parallel RC circuit since both elements experience the same generalized force ($\sigma$ or $V$) in this configuration.
For the Maxwell model, $G^{*}(\omega)/i\omega = G \tau/(1 + i \omega \tau)$, while for a parallel RC circuit $Z^{*} = C^{-1} \tau/(1 + i \omega \tau)$. Note that the prefactor for these analogous quantities is stiffness. $\tau$ is the characteristic timescale over which the system \textit{relaxes} back to equilibrium after a step in $x$ is imposed. $\tau=\eta/G$ in MS and $\tau=RC$ in IS.

Similarly, a spring and dashpot in parallel (Kelvin-Voigt model) evokes a response that resembles a capacitor and resistor in series because both elements in the circuit experience the same generalized displacement or velocity ($\dot{\gamma}$ or $\dot{Q}$). The resulting $G^*(\omega) = G (1+i\omega \tau)$ for the Kelvin-Voigt model is equivalent to $i \omega Z^* = C^{-1}(1 + i \omega \tau)$ for the series-RC circuit. \addition{Here $\tau$ describes how quickly a system reaches equilibrium following a step input in $y$ and is associated with \textit{retardation}.}

This similarity can be extended to three element models as well. The Zener model in MS, which consists of a Maxwell element \textit{in parallel} with another spring is equivalent to a parallel RC circuit \textit{in series} with another capacitor. The Debye model, which consists of a series RC element \textit{in parallel} with another capacitor is equivalent to a Kelvin-Voigt element \textit{in series} with another spring, called the Poynting-Thompson model in MS. These are illustrated in Table \ref{tab:combination_elements}. 

\addition{Overall, a series connection in MS mirrors a parallel connection in IS, and vice-versa. Accordingly, for elements in series (parallel) for IS (MS), the impedance $\Zstar$ (viscosity $\eta^*$) is additive. For the reverse combinations (parallel for IS and series for MS), the reciprocals of $\Zstar$ and $\eta^{*}$ are additive.}
This equivalence of configurations in MS and IS allows us to directly infer the time or frequency-dependent response functions of one kind of spectroscopy from the other, if the analogous response function is known.


\subsection{Conjugate Variables in Theory and Experiments}
\label{sec:variables_theory_exp}

Table \ref{tab:x_y} lists the conjugate variables -- generalized force ($y$) and displacement ($x$) -- and the velocity ($\dot{x}$). The conjugate variables in MS, $\sigma$ and $\gamma$, are also the control variables in experiments: either $\sigma$ or $\gamma$ may be perturbed, while the other is measured. In EIS, however, the conjugate variables are voltage $V$ and charge $Q$, but the experimental control variables are $V$ and current $I$, since current is more convenient to work with. On the other hand, in BDS, the conjugate variables $E$ and $D$ are used for data representation and analysis. However, $V$ and $I$ are used as control variables in experiments.
Therefore, conjugate variables in theory do not always coincide with control variables in experiments.

These distinctions are a common source of confusion. For instance, in EIS the labels `in-phase' and `out-of-phase' are applied with respect to experimental control variables ($V$ and $I$) and not conjugate variables ($V$ and $Q$). Thus, the $V$ across a resistor is in-phase with $I$, and out-of-phase with a capacitor. This can be puzzling because it is different from the convention in MS and BDS, where the labels `in-phase' and `out-of-phase' are applied with respect to theoretical conjugate variables.

\subsection{Response functions}
\label{sec:response}

\begin{table}
    \centering
    \begin{tabular}{ccc|c}
    \hline
        \textbf{MS} & \textbf{EIS} & \textbf{BDS} & \textbf{LRT} \\
        \hline 
         $G(t)$ &  & $M(t)$ & $R_x(t)$ \\
         $J(t)$ &  & $\epsilon (t)$ & $R_y(t)$ \\
         \hline
         $\eta^*\omg=\eta^\prime - i \eta^{\prime \prime}$ & $\Zstar \omg = \Zp+ i\Zpp$ & $\epsilon_0 \rho^* \omg, \, C_0 \Zstar \omg$ & $\hat{R}_x(i \omega)= \hat{R}_x^\text{cos} - i \hat{R}_x^\text{sin} $ \\
         &  &  & $\hat{R}_y(i \omega)= \hat{R}_y^\text{cos} - i \hat{R}_y^\text{sin} $ \\
         & $Y^*\omg=Y^\prime + i Y^{\prime \prime}$ & $\sigma^* \omg/\epsilon_0$ & $\hat{R}_x (i\omega)^{-1}=-\omega^2 \hat{R}_y(i\omega)$  \\
         $G^*\omg = \Gp + i \Gpp $ &   & $M^*\omg = M^\prime + i M^{\prime \prime}$ & $i\omega\hat{R}_x(i \omega)=\hat{\dot{R}}_x(i\omega)+R_x(0) $  \\
         & & & $=\omega \hat{R}_x^\text{sin} + i\omega \hat{R}_x^\text{cos}$ \\
         $J^*\omg = J^\prime - i J^{\prime \prime}$ &   & $\epsilon^* \omg = \epsilon^\prime - i \epsilon^{\prime \prime}$ &  $i\omega\hat{R}_y(i \omega)=\hat{\dot{R}}_y(i\omega)+R_y(0)$ \\
         & & & $=\omega \hat{R}_y^\text{sin} + i\omega \hat{R}_y^\text{cos}$ \\
         \hline
    \end{tabular}
    \caption{Commonly used response functions in MS, EIS and BDS and their correspondence to LRT. Here $C_0  = \epsilon_0 A/d$ is the capacitance of vacuum, $\sigma^*=i\omega \epsilon_0 \epsstar$ is the complex conductivity and $\rho^*=(\sigma^*)^{-1}$ is the the complex resistivity in BDS.}
     \label{tab:response_functions}
\end{table}

Tables \ref{tab:lrt_relations} and \ref{tab:response_functions} summarize the LRT relations and response functions commonly used in MS, EIS, and BDS.
\begin{enumerate}[label=(\roman*)]

    \item \textbf{Time-domain response}: 
    The relaxation modulus $G(t)$ and creep compliance $J(t)$ are often measured in MS by applying step jumps in $\gamma$ and $\sigma$, respectively. These time-domain functions are used in conjunction with their frequency-domain counterparts (although $G^{*}$ is reported more often than $J^{*}$). In contrast, EIS is primarily a frequency-domain technique. Time-domain response functions are explored in other techniques such as cyclic voltammetry,\cite{gosser1993cyclic} chronoamperometry,\cite{rafiee2024cyclic} and chronopotentiometry.\cite{Paleček2018} While measurements of time-domain permittivity $\epsilon(t)$ are sometimes reported for BDS, it is much more common to measure $\epsilon^{*}(\omega)$. The direct measurement of the dielectric modulus $M(t)$ is rare.
 
    \item \textbf{Frequency domain response}: There is a one-to-one correspondence between the complex shear modulus $G^*$ (compliance $J^*$) in MS and the complex dielectric modulus $ M^*$ (permittivity $\epsstar$) in BDS. These have no direct analogs in EIS, where complex impedance $\Zstar$ is the primary function of interest. $Z^*$ is analogous to complex viscosity $\eta^*$ in MS and resistivity $\rho^*$ in BDS. This is unsurprising since resistance and viscosity are conceptually similar (section \ref{Sec:elements_discussion}). $\Zstar$ is also sometimes used in BDS analysis. The prefactors $\epsilon_0$ and $C_0$ are required to balance the units according to the convention.
    
    In terms of the LRT, the complex modulus and compliance functions (in MS and BDS), which are ratios of Fourier transforms of generalized force and displacement, are special cases of $i\omega \hat{R}_x\omg$ and $i\omega \hat{R}_y\omg$, respectively. Accordingly, the terms $\omega\hat{R}_x^\text{sin}$ and $\omega\hat{R}_x^\text{cos}$ represent the component in-phase and out-of-phase with $x$, respectively. The real (imaginary) part of $i\omega \hat{R}_x$ is an even (odd) function of $\omega$ and associated with the elastic (viscous) stored (dissipated) energy. $i\omega \hat{R}_y$ is interpreted in the same manner.
    
    On the other hand, $Z^*$ maps to $\hat{R}_x\omg$ since it is a ratio of the Fourier transforms of generalized force and velocity. There is no common response function in the three spectroscopies that corresponds to $\hat{R}_y$. However, the inverse of $\hat{R}_x$  conforms to admittance $Y^*$ in EIS and conductivity $\sigma^*$ in BDS.

It is worth mentioning that $i\omega \hat{R}_y (i\omega)$ relates to the Fourier transform of $\dot{R}_y$, not $R_y$. The condition $R_y(0)=0$ indicates the absence of a free elastic element uncoupled from any dissipative element. For example, in MS, the Kelvin-Voigt model ($R_y(0)=J(0)=0$) has an elastic spring coupled to a viscous dashpot, while in the Maxwell model ($J(0) \neq 0$) they are uncoupled (see Table \ref{tab:ms_properties}).

    
\end{enumerate}

\subsection{Kramers-Kronig Relations}
\label{sec:kkr}

\begin{table}
    \centering
    \begin{tabular}{c|c|c}
    \hline
         & \multicolumn{2}{c}{\textbf{Kramer-Kronig Relations}}
         \\ \hline & &  \\ [-4ex] 
         \multirow{3}{*}{MS} &  $\displaystyle G^* (\omega) - G_e = \dfrac{i\omega}{\pi} \int_{-\infty}^\infty \dfrac{G^*(u)-G_e}{u(u-\omega)} \, du$  
         & $\displaystyle J^* \omg - J_g + \dfrac{i}{\eta_0 \omega} = \dfrac{i}{\pi} \int_{-\infty}^\infty \dfrac{J^*(u) - J_g + \dfrac{i}{\eta_0 u}}{u-\omega} du$ 
         \\ [1ex]
         & $\displaystyle G^{\prime} (\omega) - G_e= -\dfrac{2\omega^2}{\pi} \int_0^\infty \dfrac{G^{\prime \prime}(u)/u}{u^2-\omega^2} du$ 
         & $\displaystyle \Jp \omg -J_g=\dfrac{2}{\pi}\int_{0}^\infty \dfrac{u \Jpp (u) -1/\eta_0}{u^2 - \omega^2} du$
         \\ [1ex]
         & $\displaystyle G^{\prime \prime} (\omega) = \dfrac{2\omega}{\pi} \int_0^\infty \dfrac{G^{\prime}(u) - G_e}{u^2-\omega^2} du$ 
         & $\displaystyle \Jpp\omg - \dfrac{1}{\eta_0\omega} = -\dfrac{2 \omega}{\pi} \int_0^\infty \dfrac{\Jp\omg - J_g}{u^2 - \omega^2} du $
         \\ [2ex] \hline & & \\ [-3ex]
         \multirow{3}{*}{EIS} & $\displaystyle \Zstar \omg - \Zp_\infty = \dfrac{1}{i \pi} \int_{-\infty}^\infty \dfrac{\Zstar (u) - \Zstar\omg}{u-\omega} \, du $ 
         & $\displaystyle Y^* \omg - Y^\prime_\infty = \dfrac{1}{i \pi} \int_{-\infty}^\infty \dfrac{Y^* (u) - Y^*\omg}{u-\omega} \, du $
         \\ [1ex]
         & $\displaystyle  \Zp \omg -\Zp_\infty = \dfrac{2}{\pi} \int_0^\infty \dfrac{u \Zpp (u) - \omega \Zpp \omg}{u^2 - \omega^2} \, du$ 
         &  $\displaystyle  Y^\prime \omg - Y^\prime_\infty = \dfrac{2}{\pi} \int_0^\infty \dfrac{u Y^{\prime \prime} (u) - \omega Y^{\prime \prime} \omg}{u^2 - \omega^2} \, du$
         \\ [1ex]
         & $\displaystyle  \Zpp \omg  =- \dfrac{2 \omega}{\pi} \int_0^\infty \dfrac{ \Zp (u) - \Zp \omg}{u^2 - \omega^2} \, du $ 
         & $\displaystyle  Y^{\prime \prime} \omg  =- \dfrac{2 \omega}{\pi} \int_0^\infty \dfrac{ Y^{\prime} (u) - Y^{\prime \prime} \omg}{u^2 - \omega^2} \, du $
         \\ [2ex] \hline & & \\ [-3ex] 
         \multirow{3}{*}{BDS} 
         & 
          &  $\displaystyle \epsilon^* \omg - \epsilon_\infty + \dfrac{i\sigma_0}{\epsilon_0 \omega} = \dfrac{i}{\pi} \int_{-\infty}^\infty \dfrac{\epsilon^* (u) - \epsilon_\infty +  \dfrac{i\sigma_0}{\epsilon_0 u}}{u-\omega} du $ 
         \\ [1ex]
         & 
         & $\displaystyle  \epsp \omg -\epsilon_\infty = \dfrac{2}{\pi} \int_0^\infty \dfrac{u \epspp (u)}{u^2 - \omega^2} \, du $
         \\ [1ex]
         & 
         & $\displaystyle  \epspp \omg -  \dfrac{\sigma_0}{\epsilon_0 \omega} =- \dfrac{2 \omega}{\pi} \int_0^\infty \dfrac{ \epsp (u) -\epsilon_\infty}{u^2 - \omega^2} \, du$ 
         \\ [2ex]
         \hline
    \end{tabular}
    \caption{The most common Kramer-Kronig relations used in MS,\cite{Ferry1980, tschoegl2012phenomenological} EIS \cite{Lasia2014,wang2021electrochemical} and BDS.\cite{axelrod2004dielectric, kremer2002broadband} See text for explanation of symbols.}
    \label{tab:kkr}
\end{table}

The Kramers-Kronig relations (KKR) are mathematical relations that constrain the real and imaginary parts of complex response functions. They appear throughout physics wherever LRT applies. For a causal complex response function  $\chi^*\omg=\chi^\prime + i \chi^{\prime \prime}$, KKR are given by
\begin{gather}
    \chi^* \omg - \chi^\prime_\infty =  \dfrac{1}{i\pi} \int_{-\infty}^\infty \dfrac{\chi^* (u) -\chi^*(\omega)}{u-\omega} \,  du, \label{eqn:kkr_hbt1} \\
    \chi^\prime \omg - \chi^\prime_\infty = \dfrac{2}{\pi} \int_0^\infty \dfrac{u \chi^{\prime \prime} (u) - \omega \chi^{\prime \prime} \omg}{u^2 - \omega^2} \, du, \label{eqn:kkr_hbt2} \\
     \chi^{\prime \prime} \omg  = - \dfrac{2 \omega}{\pi} \int_0^\infty \dfrac{ \chi^{\prime} (u) - \chi^{\prime \prime} \omg}{u^2 - \omega^2} \, du, \label{eqn:kkr_hbt3}
\end{gather}
where $\chi^\prime_\infty=\chi^\prime(\omega \to \infty)$ is a constant. Since $\int_{-\infty}^\infty du/(u-w) = 0$, the second term ($\chi^*\omg$) may be omitted. It is retained here because it plays a useful role in numerical calculations.

For mathematically inclined readers familiar with the Hilbert transform (\rrereview{see section S3.2 in the supplementary material} for definition), it may be helpful to point out that KKR are an expression of the Hilbert transform of real causal functions. Therefore, eqn \eqref{eqn:kkr_hbt1} can be written as $\chi^*=i\hbt{\chi^*}$, where $\hbt{\cdot} $ denotes a Hilbert transform.


Table \ref{tab:kkr} summarizes the most commonly used KKR equations in MS, EIS, and BDS. In MS $G_e = G(t \to \infty)=G^*(0)$ is the equilibrium modulus, $J_g =J(0)=J^*(\omega \to \infty)$ is the glassy compliance and $1/\eta_0=\dot{J}(t \to \infty)=\lim_{\omega \to 0} i\omega J^*\omg$ represents the non-recoverable contribution to compliance. In EIS, the subscript $\infty$ stands for infinite frequency limit value, $\Zp_\infty=\Zp(\omega \to \infty)$ and $Y^\prime_\infty=Y^\prime(\omega \to \infty)$. Similarly, in BDS, $\epsilon_\infty=\epsstar(\omega \to \infty)$ and $\sigma_0$ is the DC conductance. To the best of our knowledge, KKR formulated for $M^* \omg$ have not been reported in the context of BDS.



Expressions in the Table are presented as they commonly appear in the literature. KKR for EIS directly corresponds to eqns. \eqref{eqn:kkr_hbt1} -- \eqref{eqn:kkr_hbt3}; KKR for MS and BDS can be reformulated to fit the same form. For MS $\chi^*\omg=(G^*\omg-G_e)/i\omega$ in the complex modulus representation and, $\chi^*\omg=J^{*}\omg$ with $\chi^\prime_\infty=J_g$ for complex compliance. In the latter, the $(\eta_0\omega)^{-1}$ term  is naturally eliminated when $u \chi^{\prime \prime}(u)- \omega\chi^{\prime \prime}\omg=u\Jpp(u) - \omega \Jpp \omg$ is used. Similarly, for BDS $\chi^*\omg=\epsilon^*\omg$ with $\chi^\prime_\infty=\epsilon_\infty$, where the $\sigma_0/\epsilon_0 \omega$ is automatically accounted for.


KKR are a powerful data validation tool. They can be used to assess whether perturbations are small enough to obey linearity and large enough to overcome instrument noise. In EIS, such checks of consistency are often integrated in commercial analysis software.\cite{SADKOWSKI2004241, GINERSANZ2016254, URQUIDIMACDONALD19901559} In BDS, KKR are sometimes employed, but their applicability is constrained by parasitic effects such as conduction currents and electrode polarization.\cite{kozak2017kramers, Bobrov_2010, kremer2002broadband} \review{In MS, the applicability of KKR for transformation, i.e. to estimate the imaginary component from the real component, or vice versa, is stymied by the narrow frequency window available for data collection.\cite{kkr1, kkr2, ROULEAU201366} Nevertheless, variations of Boukamp’s method which fits a generalized circuit of storage and dissipation elements,\cite{winter1997analysis, BOUKAMP199485, Boukamp_1995, Poudel2022} or the measurement model approach which analyzes the error structure of the real and imaginary components may be used when data appear questionable.\cite{agarwal1995application, Durbha1997, Singh2019}}

\subsection{Equivalent Circuits and Multimode Models}
\label{sec:multimode}

\begin{figure}
\begin{center}
\includegraphics[width=0.5\textwidth]{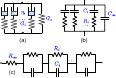}
\caption{Common equivalent circuits in (a) MS (multimode Maxwell), (b) BDS (multimode Debye), and (c) EIS (parallel RC elements in series with a resistor $R_\infty$).
 \label{fig:multimode}}
\end{center}
\end{figure}

The elementary models discussed thus far comprised different arrangements of two or three storage and dissipation elements. While such simple models are conceptually insightful, they cannot describe the rich behavior observed in real materials and systems. Multimode models are generalizations that offer a superior quantitative description by combining several building blocks in series or parallel. For example, arranging several Maxwell elements and a spring in parallel (figure \ref{fig:multimode}a) has a relaxation modulus,
\begin{equation}
G(t) = G_e + \sum_{i=1}^{n} G_i \exp(-t/\tau_i),
\label{eqn:mmm_Gt}
\end{equation}
where $\tau_i = \eta_i/G_i$ is the relaxation time of the $i$th mode. The set of modes $H_n = \{G_i, \tau_i\}$ is called the discrete relaxation spectrum. Several methods have been proposed to obtain consistent and parsimonious spectra.\cite{provencher76, baumgaertel1989determination, Bae2016,  shanbhag2020relaxation, shanbhag2019pyRespect, takeh2013computer} This representation is not unique. Indeed, it is possible to mathematically convert a multimode Maxwell model to a multimode Kelvin-Voigt model which consists of several Kelvin-Voigt elements connected in series with a \review{spring and dashpot}.\cite{emri1993generating, winter1997analysis, Loy2015, baumgaertel1989determination, shannbhag2023interconversion, elster1992using}  As $n \to \infty$, we obtain a continuous relaxation spectrum $H(\tau)$ such that,\cite{Ferry1980} 
\begin{equation}
    G(t) = G_e + \int_{-\infty}^\infty H(\tau)\, e^{-t/\tau} d \ln\tau.  \label{eqn:ms_CRS}
\end{equation}
which, unlike $H_n$, is unique in principle. \review{Thus, different multimode models or $H_n$ represent different approximations of $H(\tau)$.} In practice, nonuniqueness does not hamper common uses of discrete relaxation spectra such as interconversion to obtain any time- or frequency-dependent response function.\cite{shanbhag2025does} 

Analogous to the multimode Maxwell model, we can define a multimode RC circuit (figure \ref{fig:multimode}c) in which several parallel RC elements are strung together in series with an additional resistor. The resulting impedance
\begin{equation}
    \Zstar \omg= R_\infty + \sum_{i=1}^{n} \dfrac{R_i}{1+i \omega \tau_i}, \label{eqn:zstar_drs}
\end{equation}
with $\tau_i = R_i C_i$ describes a relaxation spectrum. As $n \rightarrow \infty$, we obtain a continuous spectrum or distribution of relaxation times,\cite{ZHANG2015464, ivers2017evaluation} which can be used to deconvolve poorly resolved semicircles in a Nyquist plot. 

Similarly, a broad dielectric loss spectrum can be described as a superposition of multiple Debye units  (figure \ref{fig:multimode}b) or with a continuous distribution of relaxation or retardation times $L(\tau)$:
\begin{align}
    \epsstar \omg & = \epsilon_{\infty} + \Delta \epsilon \int \dfrac{L(\tau)}{1+i\omega\tau}\, d\ln \tau\\
    \epsilon(t)  & =  \epsilon_{\infty} + \Delta \epsilon \int L(\tau) \left[ 1 - \exp\!\left(-\dfrac{t}{\tau}\right) \right] \, d\ln \tau.
    \label{eqn:bds_crs}
\end{align}

Multimode models and their continuous analogs are used for characterization and interconversion. They are also used for data validation: the inability to describe response functions with discrete or continuous spectra is often a sign that KKR are violated, and that nonlinear effects may be contaminating measurements.\cite{Lasia2014, agarwal1995application} This flexible fitting approach is often not preferred in EIS because it provides little mechanistic insights. Instead the primary objective is to identify circuit elements that resolve distinct electrochemical processes. Nevertheless, multimode models are often used for data validation, noise reduction, resolving closely spaced relaxation timescales, and indirectly determining equivalent circuits.\cite{ivers2017evaluation} In MS and BDS, specialized models such as the stretched exponential or Havriliak-Negami equations and fractional models are often used because they allow us to represent system behavior more parsimoniously.


\section{Summary and Conclusions}

Spectroscopic techniques share a common foundation in linear response theory, despite differences in applications and instrumentation. This work highlights the conceptual similarities and differences in conventions between three different spectroscopies: MS, EIS, and BDS. 

Conjugate variables in the LRT may be thought of as a generalized force and a generalized displacement. These variables take different forms in MS, EIS and BDS to LRT as summarized in Table \ref{tab:x_y}. LRT tells us how small perturbations in one variable affects its conjugate.  This is mediated via complex response functions that can be thought of as generalized modulus, compliance, or impedance (section \ref{sec:response}). Importantly, experimental control variables may (MS) or may not (EIS and BDS) coincide with the conjugate variables as discussed in section \ref{sec:variables_theory_exp}. 



The linear response of materials or systems can be modeled as an equivalent circuit of building blocks that describe energy storage (spring or capacitor) and energy loss (dashpot or resistor). While such circuits are understood to be idealizations in MS and BDS, they are taken more seriously in EIS analysis where circuit elements model specific electrochemical processes. Thus, the evaluation of equivalent circuits can be automated for MS and BDS, but requires significant domain knowledge for EIS. Equivalent circuits are helpful for interconversion between response functions, and for data validation via Kramer-Kronig relations.

Overall, this perspective offers a consistent approach for a conceptual comparison among the spectroscopic methods discussed here. This framework can also be used to establish analogies with other forms of spectroscopy. This enables a cross-pollination of ideas extending beyond the LRT framework, including data analysis protocols such as elemental  and empirical models, data validation tools such as Kramers-Kronig relations, and extensions to nonlinear analysis.

\section*{Supporting Information}
\rrereview{Supporting Information: An overview of (i) the Laplace and Fourier transforms, their interrelationship, as well as (ii) the convolution integral and Fourier transforms for real causal functions}






\section*{Acknowledgments}

SS acknowledges support from the Department of Scientific Computing at Florida State University. SM and YMJ acknowledge financial support from the Science and Engineering Research Board, Government of India (Grants CRG/2022/004868 and JCB/2022/000040)  .

\clearpage

\section*{List of Symbols}

\begin{table}[ht]
   \centering
   \begin{tabular}{l l r}
        \hline
        \textbf{Symbol} & \textbf{Description} \\
        \hline
        $\sigma$ & Shear Stress \\
        $\gamma$ & Shear strain \\
        $\dot{\gamma} = d\gamma/dt$ & Shear rate \\
        $G(t)$ & Shear modulus \\
        $J(t)$ & Shear creep compliance \\
        $G_0$ & Modulus of spring  \\
        $\eta_0$ & Viscocity of dashpot\\
        $G^*$ & Complex modulus \\
        $J^*$ & Complex compliance \\
        $\eta^*$ & Complex viscocity \\
        $\Gp (\Jp)$ & Storage modulus (compliance) \\
        $\Gpp (\Jpp)$ & Loss modulus (compliance) \\
        $\gamma_0 (\sigma_0)$ & Oscillatory strain (stress) amplitude \\
        $\tau=\eta_0/G_0$ & Relaxation time \\
        $H(\tau)$ & Relaxation time spectra \\
        \hline
   \end{tabular}
   \caption{List of symbols used in mechanical spectroscopy}
   \label{tab:symbols_ms}
\end{table}

\begin{table}[ht]
   \centering
   \begin{tabular}{l l r}
        \hline
        \textbf{Symbol} & \textbf{Description}  \\
        \hline
        $V$ & Voltage \\
        $Q$ & Charge \\
        $I=dQ/dt$ & Current  \\
        $C$ & Capacitance \\
        $R$ & Resistance  \\
        $L$ & Inductance \\
        $\Zstar $ & Complex Impedance \\
        $Y^*$ & Complex Admittance \\
        $\Zp$ & Resistance \\
        $\Zpp$ & Reactance \\
        $Y^{\prime}$ & Conductance \\
        $Y^{\prime \prime}$ & Susceptance \\
        $V_0 (I_0)$  & Oscillatory voltage (current) amplitude \\
        $\delta \omg = \tan^{-1} (\Zpp/\Zp)$ & Phase shift \\
        $\Zstar_R, \Zstar_C, \Zstar_L$ & Impedance of resistor, capacitor and inductor \\
        $\tau=RC$ & Characteristic timescale \\
        
        \hline
   \end{tabular}
   \caption{List of symbols used in electrochemical impedance spectroscopy}
   \label{tab:symbols_eis}
\end{table}

\begin{table}[ht]
   \centering
   \begin{tabular}{l l r}
        \hline
        \textbf{Symbol} & \textbf{Description}  \\
        \hline
        $E$ & Electric field \\
        $D$ & Dielectric displacement \\
        $j=dD/dt$ & Current density \\
        $\epsilon_0$ & Vacuum permittivity \\
        $C_0 = \epsilon_0 A/d$ & Vacuum capacitance \\
        $\epsilon$ & Relative permittivity \\
        $\phi (t)$ & Relaxation function in time domain  \\
        $\epsilon_\infty=\epsilon(t \to\infty)$ & High frequency permittivity  \\
        $\epsilon_s = \epsilon(t \to 0)$ & Static or low frequency permittivity  \\
        $\Delta \epsilon = \epsilon_s-\epsilon_\infty$ & Dielectric strength \\
        $\epsilon (t)$ & Dielectric permittivity in time domain  \\
        $M(t)$ & Dielectric modulus in time-domain \\
        $\epsstar$ & Complex permittivity \\
        $M^*$ & Complex dielectric modulus \\
        $\sigma^* (\rho^*)$ & Complex conductivity (resistivity) \\
        $\epsp, \epspp$ & Storage and loss permittivity \\
        $M^{\prime}, M^{\prime \prime}$ & Storage and loss permittivity \\
        $\sigma_\text{dc}$ & DC conductivity \\
        $\tau = \epsilon RC_0$ & Relaxation time \\
        \hline
   \end{tabular}
   \caption{List of symbols used in broadband dielectric spectroscopy}
   \label{tab:symbols_bds}
\end{table}

\clearpage

\clearpage

\printbibliography



\end{document}


\begin{center}
\textbf{\huge{Supplementary Material}}\\
\vspace{0.5cm}
\textbf{When Spectroscopies Speak the Same Language: Unifying Rheology, Electrochemical Impedance, and Dielectrics}\\
\vspace{0.5cm}

\medskip

Shivangi Mittal$^{1}$, Sachin Shanbhag$^{2,*}$, and Yogesh M. Joshi$^{1,*}$\\ 

${}^{1}$Department of Chemical Engineering, Indian Institute of Technology
Kanpur, India\\
${}^{2}$Department of Scientific Computing, Florida State University, Tallahasee, Florida, USA

\end{center}

\vspace{0.5cm}

\tableofcontents

\clearpage

\section{Laplace Transform}
\label{sec:lt}

The Laplace transform of a function $f(t)$ is defined as
\begin{equation}
\lt{f(t)} = \tilde{f}(s) = \int_{0}^{\infty} f(t)\, e^{-st}\, dt.
\label{eqn:lt}
\end{equation}
For the Laplace transform to exist $f(t)$ must be piecewise continuous and of exponential order, i.e. $|f(t)| < M e^{at}$ for some constants $M$ and $a$.\cite{kreyszig2007advanced} We list selected properties of Laplace transforms that are useful in the context of the LRT:
\begin{enumerate}
\item \textit{Derivatives}: The Laplace transform of the derivative $\dot{f} = df/dt$ is related to the Laplace transform of $f(t)$ via
\begin{equation}
\lt{\dot{f}(t)} = s\tilde{f}(s) - f(0).
\label{eqn:lt_derivative}
\end{equation}

Since linear response functions obey time-invariance it is common to define $t = 0$, without any loss of generality, by stipulating that the system is at equilibrium for $t < 0$. Therefore, by convention, we take the equilibrium values  $x(t = 0^{-}) = x_\text{eq} = 0$ and $y(t = 0^{-}) = y_\text{eq} = 0$. However, it should be noted that though $R_x(t = 0^{-})=R_y(t = 0^{-})=0$, $R_x(t=0) \neq 0$  and $R_y(t=0) \neq 0$ usually holds. 
This implies, for example, that $\lt{\dot{R}_{x}(t)} = s\tilde{R}_{x}(s)-R_x(0)$ and $\lt{\dot{x}(t)} = s\tilde{x}(s)$.

\item \textit{Convolution Theorem}: The Laplace transform of a convolution is the product of Laplace transforms of the constituent functions. That is 
\begin{equation}
\lt{(f * g)(t)} = \tilde{f}(s)\, \tilde{g}(s).
\label{eqn:lt_convolution}
\end{equation}

\end{enumerate}


\section{Fourier Transform}
\label{sec:ft}

We define the complex Fourier transform of a function $f(t)$ as
\begin{equation}
\ft{f(t)} = \hat{f}(\omega) = \int_{-\infty}^{\infty} f(t)\, e^{-i \omega t} dt.
\label{eqn:ft}
\end{equation}
For the Fourier transform to exist, $f(t)$ has to be absolutely integrable; i.e. $\int_{-\infty}^{\infty} |f(t)|\, dt < \infty$. The units of $\hat{f}(\omega)$ are not the same as the units of $f(t)$. They differ by units of time (assuming that $t$ stands for time). There are other definitions of the Fourier transform which use the opposite sign in front of the exponential ($e^{+i \omega t}$ instead of $e^{-i \omega t}$) and different pre-factors in front of the integral ($2\pi$ or $\sqrt{2\pi}$ instead of 1). This can be a source of confusion. With our convention, the inverse Fourier transform is defined as
\begin{equation}
f(t) = \dfrac{1}{2\pi}  \int_{-\infty}^{\infty} \hat{f}(\omega) \, e^{i \omega t} d\omega.
\label{eqn:ift}
\end{equation}

Building upon eqn. \eqref{eqn:ft} with $e^{-i \omega t}=\cos{\omega t} - i \sin{\omega t}$, the sine and cosine transforms of the periodic function $f(t)$ can be respectively defined as,
\begin{gather}
    \hat{f}^\text{sin}(\omega) = \int_{-\infty}^{\infty} f(t)\, \sin{\omega t} \, d\omega, \\
    \hat{f}^\text{cos}(\omega) = \int_{-\infty}^{\infty} f(t)\, \cos{\omega t} \, d\omega.
\end{gather}
Here $\hat{f}^\text{sin}(\omega)$ and $\hat{f}^\text{cos}(\omega)$ account for the odd and even components of the function, respectively. This yields,
\begin{equation}
    \hat{f}(\omega)= \hat{f}^\text{cos}(\omega) - i \hat{f}^\text{sin}(\omega).
\end{equation}

\subsection{Relationship with Laplace transform}

A Laplace transform is a generalization of a \textit{one-sided} Fourier transform. If $f(t)$ is causal or one sided (so that $f(t < 0) = 0$), and absolutely integrable,\cite{kreyszig2007advanced}
\begin{equation}
\hat{f}(\omega) = \lim_{s \rightarrow i\omega} \tilde{f}(s).
\label{eqn:lt_and_ft}
\end{equation}
For example, consider $f(t) = e^{-t} H(t)$, where $H(\cdot)$ is the Heaviside step function. The Laplace transform is $\tilde{f}(s) = 1/(s+1)$ and according to eqn \eqref{eqn:lt_and_ft}, $\hat{f}(\omega) = (1+i\omega)^{-1}$. However, there are functions for which the Laplace transform exists, but Fourier transform does not. For example, $f(t) = t H(t)$. Here, $\tilde{f}(s) = s^{-2}$, while $\hat{f}(\omega)$ does not exist because $f(t)$ is not absolutely integrable.

\medskip

When the Fourier transform exists, it inherits the corresponding properties of the Laplace transform. In particular,
\begin{align}
\hat{\dot{f}}(\omega) & = i \omega \hat{f}(\omega) - f(0) \\
\ft{f*g} & = \hat{f}(\omega)\hat{g} (\omega).
\end{align}

\section{Real Causal Functions}

\subsection{Convolution Integral}
\label{sec:convolution}

A note on the limits of the convolution integral might be helpful to some readers, since most proofs of this theorem assume that either one or both $f$ and $g$ are causal functions, i.e. $f(t < 0) = g(t < 0) = 0$. Unlike the second to last term in eqn \eqref{m-eqn:lrt-conv} of the manuscript where the domain of integration is infinite $(-\infty, \infty)$, a convolution is defined with a semi-infinite $[0, \infty)$ domain of integration when either $f$ or $g$ is a causal function.  
\begin{equation}
(f * g)(t) \equiv \int_{0}^{\infty} f(t)\, g(t - \tp)\, d\tp.
\end{equation}
This is the case for eqn. \eqref{m-eqn:lrt-conv} since response functions are causal. Similarly, when both $f$ and $g$ are causal the integration is over a finite domain $[0, t]$,
\begin{equation}
(f * g)(t) \equiv \int_{0}^{t} f(t)\, g(t - \tp)\, d\tp.
\end{equation}
as seen in eqn. \eqref{m-eqn:convolution_time}.




\subsection{Fourier transforms}
\label{sec:real_causal_func}

The Fourier transform of a real causal function $\chi(t)$ is
\begin{align}
\hat{\chi}(\omega) & = \int_{-\infty}^{\infty} \chi(t) e^{-i \omega t} dt 
= \int_{0}^{\infty} \chi(t) \left[ \cos \omega t - i \sin \omega t \right]\, dt \notag\\
& = \hat{\chi}^\text{cos}(\omega) - i \hat{\chi}^\text{sin}(\omega)
\label{eqn:real_causal}
\end{align}
where the last line defines the sine ($\hat{\chi}^\text{sin}(\omega)$) and cosine ($\hat{\chi}^\text{cos}(\omega)$) transforms of a real causal function.
\begin{align}
\hat{\chi}^\text{cos}(\omega) & = \int_{0}^{\infty} \chi(t) \cos \omega t \, dt \notag \\
\hat{\chi}^\text{sin}(\omega) & = \int_{0}^{\infty} \chi(t) \sin \omega t \, dt.
\label{eqn:sine_cosine_transforms}
\end{align}
Be aware that it is also common in the literature to define the sine transform of $\chi(t)$ with a minus sign (so that $\hat{\chi} = \hat{\chi}^\text{cos}(\omega) + i \hat{\chi}^\text{sin}(\omega)$ instead of $\hat{\chi} = \hat{\chi}^\text{cos}(\omega) - i \hat{\chi}^\text{sin}(\omega)$). Regardless, since $\chi(t)$ can be decomposed into even and odd parts, $\hat{\chi}^\text{cos}(\omega)$ is an even function, while $\hat{\chi}^\text{sin}(\omega)$ is an odd function of $\omega$. 

Furthermore, the real and imaginary parts are constrained via a Hilbert transform, 
\begin{equation}
    \hat{\chi}^\text{cos} \omg = \hbt{\hat{\chi}^\text{sin}} \quad \text{and} \quad \hat{\chi}^\text{sin} \omg = -\hbt{\hat{\chi}^\text{cos}}, \label{eqn:kkr_ht_sine_cos}
\end{equation}
where, the Hilbert transform is defined as a principal value integral, 
\begin{equation}
    \hbt{\hat{\chi}}\omg = \dfrac{1}{\pi} \int_{-\infty}^\infty \dfrac{\hat{\chi} (u)}{u-\omega} du. \label{eqn:ht}
\end{equation}
Eqn. \eqref{eqn:kkr_ht_sine_cos} can be compactly written as,
\begin{equation}
    \hat{\chi} = i \hbt{\hat{\chi}}. \label{eqn:kkr_ht}
\end{equation}
which constitutes the Kramer-Kronig relations (KKR), 
\begin{equation}
\hat{\chi}(\omega) = \dfrac{i}{\pi} \int_{-\infty}^{\infty} \dfrac{\hat{\chi}(u)}{u - \omega}\, d\omega, 
\label{eqn:kkr_general}
\end{equation}
The prefactor of $i$ is responsible for the connection between the real ($\hat{\chi}^\text{cos}$) and imaginary (-$\hat{\chi}^\text{sin}$) parts, and can be explicitly expressed as,
\begin{align}
\hat{\chi}^\text{cos}(\omega) & = \dfrac{2}{\pi} \int_{0}^{\infty} \dfrac{u \hat{\chi}^\text{sin}(u)}{u^2 - \omega^2} du \notag \\
\hat{\chi}^\text{sin}(\omega) & = \dfrac{2\omega}{\pi} \int_{0}^{\infty} \dfrac{\hat{\chi}^\text{cos}(u)}{\omega^2 - u^2} du.
\label{eqn:kkr_sine_cosine}
\end{align}
KKR holds only when the complex function obeys (a) analyticity in the upper half of the complex plane, (b) Hermitian symmetry $\hat{\chi}\omg=\hat{\chi}(-\omega)$, which implies $\hat{\chi}^\text{cos} \omg=\hat{\chi}^\text{cos} (-\omega)$ and $\hat{\chi}^\text{sin} \omg=-\hat{\chi}^\text{sin} (-\omega)$, (c) finiteness requiring $\hat{\chi}(\omega \to 0)$ and $\hat{\chi} (\omega \to \infty)$ to be finite, and (d) fast decay at large $\omega$, typically faster than $\hat{\chi}\sim \omega^{-1}$.



\printbibliography